\documentclass[aps,prb,twocolumn,amsmath,amssymb,superscriptaddress,floatfix]{revtex4}
\usepackage{graphicx}
\usepackage{graphics}
\usepackage{bm}
\usepackage[usenames]{color}
\usepackage{colordvi}
\usepackage{units}
\usepackage{bbm}
\usepackage{booktabs}
\usepackage{array}
\usepackage{placeins}
\usepackage{epsfig}
\usepackage{lscape}
\usepackage{changes}
\usepackage{braket}
\usepackage{tabularx}
\newcolumntype{L}[1]{>{\raggedright\arraybackslash}p{#1}} 
\newcolumntype{C}[1]{>{\centering\arraybackslash}p{#1}} 
\newcolumntype{R}[1]{>{\raggedleft\arraybackslash}p{#1}} 
\usepackage[colorlinks=true,linkcolor=blue,citecolor=blue,urlcolor=blue]{hyperref}
\newcommand{\be}{\begin{equation}}
\newcommand{\ee}{\end{equation}}
\newcommand{\beqn}{\begin{eqnarray}}
\newcommand{\eeqn}{\end{eqnarray}}

\definecolor{mymagenta}{rgb}{1.0,0.0,1.0}
\definecolor{mycyan}{rgb}{0.0,1.0,1.0}
\definecolor{myyellow}{rgb}{1.0,1.0,0.0}
\definecolor{myorange}{rgb}{1.0,0.27,0.0}

\definecolor{dark-gray}{HTML}{a0a0a0}
\definecolor{dark-red}{HTML}{8b0000}
\definecolor{dark-green}{HTML}{006400}
\definecolor{dark-blue}{HTML}{00008b}
\definecolor{gold}{rgb}{1.0,0.84,0.0}
\definecolor{dark-turquoise}{HTML}{00ced1}

\bibstyle{apsrev.bib}

\begin{document}

\title{Extreme statistics of the excitations in the random transverse Ising chain}
\author{Istv\'an A. Kov\'acs}
\email{istvan.kovacs@northwestern.edu}
\affiliation{Department of Physics and Astronomy, Northwestern University, Evanston, IL 60208-3112, USA}
\author{Tam\'as Pet\H {o}}
\email{petotamas0@gmail.com}
\affiliation{Institute of Theoretical Physics, Szeged University, H-6720 Szeged, Hungary}
\author{Ferenc Igl{\'o}i}
\email{igloi.ferenc@wigner.hu}
\affiliation{Wigner Research Centre for Physics, Institute for Solid State Physics and Optics, H-1525 Budapest, P.O. Box 49, Hungary}
\affiliation{Institute of Theoretical Physics, Szeged University, H-6720 Szeged, Hungary}
\date{\today}

\begin{abstract}
In random quantum magnets, like the random transverse Ising chain, the low energy excitations are localized in rare regions and there are only weak correlations between them. It is a fascinating question whether these correlations are completely irrelevant in the sense of the renormalization group. To answer this question, we calculate
the distribution of the excitation energy of the random transverse Ising chain in the disordered Griffiths phase with high numerical precision by the strong disorder renormalization group method and - for shorter chains - by free-fermion techniques. Asymptotically, the two methods give identical results, which are well fitted by the Fr\'echet limit law of the extremes of independent and identically distributed random numbers. Given the finite size corrections, the two numerical methods give very similar results, but they differ from the correction term for uncorrelated random variables. This fact shows that the weak correlations between low-energy excitations in random quantum magnets are not entirely irrelevant.
\end{abstract}


\maketitle

\section{Introduction}
\label{sec:intr}

Many-body systems in the presence of quenched disorder have unusual dynamical properties due to rare-region effects. In these systems - due to extreme fluctuations of strong couplings - domains are formed, which can remain locally ordered even in the paramagnetic phase. The relaxation time, $\tau$, associated with turning the spins in such domains can be extremely large and it has no upper limit in the thermodynamic limit. This type of Griffiths singularities are responsible for non-analytical behaviour of several physical quantities (susceptibility, specific heat, auto-correlation function) in the so called Griffiths phase, which is an extended part of the paramagnetic phase\cite{griffiths}.

In random quantum systems Griffiths-effects are stronger than in classical ones, which is manifested in power-law decay of the auto-correlation function, as well as power-law singularities of the susceptibility and the specific heat at low temperatures\cite{mccoy,vojta}. In random quantum magnets with discrete symmetry, such as in the random transverse Ising model (RTIM) the low-energy excitations are localised and their properties can be successfully studied by the so called strong disorder renormalization group (SDRG) method\cite{im}. As initiated be Ma, Dasgupta and Hu\cite{mdh} and further developed by Fisher\cite{fisher} the SDRG is a local renormalization method, in which quantum and disorder fluctuations are treated at the same time and degrees of freedom with a large excitation energy are successively eliminated. In these random quantum magnets the SDRG method is expected to provide asymptotically exact results not only at the critical point, the properties of which are governed by an infinite disorder fixed point\cite{daniel_review}, but in the Griffiths phase as well, at least for the dynamical singularities.

In phenomenological descriptions it is often assumed that the localized excitations in random quantum magnets are independent\cite{thill_huse}. For example distribution of low-energy excitations in these systems are well approximated by the Fr\'echet distribution\cite{jli}, which represents the limit law of the extremes of independent and identically distributed (iid) random numbers. Recently extreme value statistics (EVS) has been applied to several problems, we can mention earthquakes, tsunamis, extreme flooding, big wildfire, extremes of climate, stock market risks in finance, sport records, etc\cite{fisher_tippett,gumbel,gnedenko,weibull,galambos,ev}. In a mathematical point of view complete understanding of extreme value statistics is known for uncorrelated random numbers, in which case also the convergence to the limit laws is derived by mathematical\cite{haan} and renormalization group\cite{prl,pre,bertin,zarfaty} (RG) methods. The iid limit distributions also apply to weakly correlated random numbers, while new types of limit distributions appear for the strongly correlated cases\cite{mps}.

It is an intriguing problem to what extent the localised excitations in random quantum magnets are independent? Whether the weak correlations between the rare regions are completely irrelevant or these are manifested in some effects, such as in the form of finite-size corrections? In this paper we make an effort to answer this question and consider the paradigmatic model, the RTIM in one dimension (1D) and calculate numerically the distribution of the first energy gap by the asymptotically exact SDRG method with high accuracy. For moderate $L$ system sizes, we use also free-fermion techniques\cite{JW,lieb61,pfeuty79} to calculate the gaps of the random samples exactly. The distribution of the gaps for finite $L$ are compared with an appropriate Fr\'echet distribution and their difference is analysed through finite-size scaling. To check the potential form of numerical errors the same type of numerical test is repeated for iid random numbers, too.

The rest of the paper is organised in the following way. In Sec.\ref{sec:model} the RTIM model is introduced and in Sec.\ref{sec:method} the methods to calculate its excitation energy are described. In Sec.\ref{sec:results} the distributions of gaps of the RTIM are calculated and the finite-size corrections are compared with the analytical results of uncorrelated variables. To test the numerical accuracy we repeat this analysis for uncorrelated Kesten variables\cite{kesten73}. Our paper concludes with a discussion in the final section. Related results about the distribution of extremes of uncorrelated variables are recapitulated in the Appendix.

\section{Model and known results}
\label{sec:model}

Here, we consider the RTIM in 1D defined by the Hamiltonian
\be
\hat{H}=-\sum_{i=1}^{L-1} J_i \sigma_{i}^{z} \sigma_{i+1}^{z}
-\sum_{i=1}^L h_i \sigma^x_{i}\;,
\label{Hamilton}
\ee
in terms of the $\sigma_{i}^{x,z}$ Pauli matrices at site $i$ and the nearest neighbour couplings, $J_i>0$ and the transverse fields, $h_i>0$, are taken from the distributions, $\pi_1(J)$ and $\pi_2(h)$, respectively. Generally we use open boundary conditions (OBC-s) and work at zero temperature, $T=0$.

In the thermodynamic limit, $L \to \infty$, the control-parameter is defined as\cite{fisher}:
\be
\delta=\frac{[\ln h]_{\rm av}-[\ln J]_{\rm av}}{\rm{var}(h)+\rm{var}(J)}\;,
\ee
where $[\cdots]_{\rm av}$ denotes averaging over quenched disorder and $\rm{var}(x)$ stands for the variance of $x$. For $\delta<0$ ($\delta>0$) the system is in the ordered ferromagnetic (disordered paramagnetic) phase and at $\delta=0$ there is a random quantum critical point. According to SDRG calculations\cite{fisher,ijl01,Igloi} and numerical results\cite{young_rieger,bigpaper} the critical behaviour of the system is controlled by an infinite disorder fixed point\cite{daniel_review}. For example, the energy scale, defined by the lowest gap, $\varepsilon$, and the length-scale are related as
\be
\ln \varepsilon \sim L^{1/2},\quad \delta=0\;.
\ee
In the paramagnetic phase this relation is in a power-law form
\be
\varepsilon \sim L^{-z},\quad \delta>0\;,
\label{L_z}
\ee
which is due to Griffiths singularities. Here, the dynamical exponent, $z=z(\delta)$, depends on the distance from the critical point and is given by the positive root of the equation~\cite{Z,ijl01,Igloi}:
\be
      \left[\left(\frac{J}{h} \right)^{1/z}\right]_{\rm av}=1\;,
      \label{eq:z_eq}
\ee
which in the vicinity of the critical point diverges as: $z \approx 1/(2\delta)$.

The distribution of the first gap has been calculated analytically through the SDRG method\cite{fisher_young}:
\be
P_L(\varepsilon;z)=\frac{1}{z}u^{1/z-1}\exp\left(-u^{1/z}\right)\;,
\label{P_L_SDRG}
\ee
in terms of $u=u_0 \varepsilon (L/\xi)^z$, $\xi$ being the correlation length and $u_0$ is a constant, defined by the standardisation. This relation is valid for $L \gg \xi$ and for $\delta \ll 1$, in which limit $\xi \sim \delta^{-2}$. We observe that Eq.(\ref{P_L_SDRG}) is just the Fr\'echet distribution, see in Eq.(\ref{eq:M_Fr_v}).

An approximate form of the distribution of the low-energy excitations can be obtained from the assumption that these excitations are localised and are due to extreme fluctuations of say $n \ll L$ consecutive strong bonds\cite{thill_huse}. The probability to find such a strongly connected cluster in the system is given by: $P_L(n) \sim L \exp(-an)$. At the same time the excitation energy due to such a cluster is exponentially small: $\varepsilon \sim \exp(-bn)$. Combining these expressions we obtain $P_L(\ln \varepsilon) \sim L\varepsilon^{1/z}$, with $z=b/a$ being the dynamical exponent in agreement with Eq.(\ref{L_z}). Then, the cumulated distribution of the relaxation times, $\tau \sim 1/ \varepsilon$ is obtained in this approach:
\be
\mu(\tau) \approx 1-A \tau^{-1/z},\quad \tau \gg 1\;.
\label{M_tau}
\ee
If we assume that the excitations are uncorrelated following the reasoning in the Appendix we arrive at the distribution in Eq.(\ref{P_L_SDRG}), which is calculated by the SDRG approach in the given limits. 

In this paper, we aim to study this problem in more detail consider the following points. i) To calculate the distribution function through numerical iteration of the SDRG approach and to check if the result in Eq.(\ref{P_L_SDRG}) is valid in the entire Griffiths phase. ii) To confirm the results of the SDRG calculation with the exact gaps obtained through free-fermion techniques. iii) To check the form of the finite-size corrections of the two methods and to compare with the analytical result for the extremes of uncorrelated random variables.

\section{Methods to calculate the gap in the RTIM}
\label{sec:method}

The energy scale of the model is given by the lowest excitation energy of the Hamiltonian in Eq.(\ref{Hamilton}). This is calculated for finite chains by the asymptotically exact SDRG method through iteration and for shorter chains by free-fermion techniques.

\subsection{The SDRG method}
\label{sec:SDRG}

In the SDRG procedure\cite{im} we perform consecutive decimation steps, each time considering the local excitations, say at position $i$. These excitations correspond to either couplings or sites, having the value of the associated gaps: $2J_i$ and $2h_i$, respectively. These gaps are sorted in descending order and the largest one, denoted by $\Omega$, which sets the energy-scale in the problem, is eliminated. Then, between remaining degrees of freedom, new terms in the Hamiltonian are generated through perturbation calculation. This procedure is successively iterated, during which $\Omega$ monotonously decreases. At the fixed point, with $\Omega^*=0$,
one makes an analysis of the distribution of the different parameters and calculates the scaling properties. In the following, we describe the elementary decimation steps.

\subsubsection{Strong-coupling decimation}

In this case, the largest local term in the Hamiltonian is a coupling, say $\Omega=J_i$, connecting sites $i$ and $i+1$ and  the two-site Hamiltonian is given by
\begin{equation}
\hat{H}_{cp}=-J_i \sigma^z_{i}\sigma^z_{i+1}-h_i\sigma^x_{i} -h_{i+1}\sigma^x_{i+1}\;.
\label{strong-coupling}
\end{equation}
The spectrum of $\hat{H}_{cp}$ contains four levels, the lower two being separated from higher two by a gap of $2J_i$. We omit the higher two levels, corresponding to merging the two strongly coupled sites into a spin cluster in the presence of a (renormalized) transverse field $\tilde{h}$, the value of which is given by second-order perturbation calculation
\be
\tilde{h}=\frac{h_i h_{i+1}}{J_i}\;.
\ee

\subsubsection{Strong-transverse-field decimation}

In this case, the largest local term is a transverse field, say $h_i$, and due to its large value this site does not contribute to the longitudinal magnetisation and therefore it is eliminated.
The renormalised coupling between the remaining sites is given by
\begin{equation}
\tilde{J}=\frac{J_{i-1}J_i}{h_i}\;,
\end{equation}
which is calculated by second order perturbation method.

To calculate the smallest gap of a given sample, $\varepsilon$, we perform $(L-1)$ decimation steps up to the last spin cluster having an effective transverse field $\tilde{h}=\varepsilon/2$.

\subsection{Free-fermion technique}
\label{sec:Free_fermion}

In this method, $\hat{H}$ is expressed in terms of
spinless free fermions \cite{JW,lieb61,pfeuty79}. In the first step the spin operators
$\sigma_i^{x,y,z}$ are mapped to fermion creation (annihilation) operators
$c_i^\dagger$ ($c_i$) by using the Jordan-Wigner
transformation \cite{JW}:  $c^\dagger_i=a_i^+\exp\left[\pi \imath \sum_{j=1}^{i-1}a_j^+a_j^-\right]$
and $c_i=\exp\left[\pi \imath
\sum_{j=1}^{i-1}a_j^+a_j^-\right]a_i^-$, where $a_j^{\pm}=(\sigma_j^x \pm \imath\sigma_j^y)/2$, 
and the Ising Hamiltonian in Eq.(\ref{Hamilton}) is written in a quadratic form
\beqn
{\hat{H}}=
-\sum_{i=1}^{L}h_i \left( 2 c^\dagger_i c_i-1 \right) -
\sum_{i=1}^{L-1} J_i(c^\dagger_i-c_i)(c^\dagger_{i+1}+c_{i+1})\;.
\label{ferm_I}
\eeqn
In the second step, the Hamiltonian in Eq.(\ref{ferm_I}) is diagonalized through a canonical transformation \cite{lieb61}, 
in terms of the new fermion creation (annihilation) operators $\eta_k^\dag$ ($\eta_k$)
\be
{\hat{H}}=\sum_{k=1}^L \epsilon_k \left( \eta_k^{\dag} \eta_k -\frac{1}{2}\right) \;.
\label{H_free}
\ee
The energies of free fermionic modes, $\epsilon_k$, are given by the eigenvalues of a $2L \times 2L$ tridiagonal matrix
\be
T =
\begin{pmatrix}
 0  & h_1 &     &       &       &       &     \cr
h_1 &  0  & J_1 &       &       &       &     \cr
 0  & J_1 &  0  & h_2   &       &       &     \cr
    &     & h_2 &  0    &\ddots &       &     \cr
    &     &      &\ddots&\ddots &J_{L-1}&     \cr
    &     &      &      &J_{L-1}&   0   & h_L \cr
    &     &      &      &       &  h_L  &  0  \cr
\end{pmatrix}
\ee
and we consider only the $\epsilon_k \ge 0$ part of the spectrum\cite{igloiturban96}. The smallest gap of ${\hat{H}}$ in Eq.(\ref{Hamilton}) is given by $\varepsilon={\rm min}|\epsilon_k|$.

\section{Numerical results}
\label{sec:results}
In the numerical calculation, the parameters of the Hamiltonian, $J_j$ and $h_j$ are taken 
from box-like distributions
\begin{align}
  \begin{split}
    \pi_1(J) &=
    \begin{cases}
      1 & \hspace*{0.65cm}\text{for } 0<J\le J_0\,,\\
      0 & \hspace*{0.65cm}\text{otherwise.}
    \end{cases} \\
    \pi_2(h) &= 
    \begin{cases}
      1/h_0 & \text{for } 0 < h \le h_0\,,\\
      0 & \text{otherwise,} 
    \end{cases}
  \end{split} 
  \label{eq:J_distrib} 
\end{align}
and set the energy-scale with $J_0=1$. In this case the critical point of the RTIM is located at $h_0=1$ and in the disordered phase, $h_0>1$, the dynamical exponent satisfies the equation:
\be
(1-z^{-2})h_0^{1/z}=1\;,
\label{eq:z_box}
\ee
see in Eq.(\ref{eq:z_eq}).
In the vicinity of the critical point $z$ diverges as $z \approx \tfrac{1}{\ln h_0}$, $h_0 \to 1^{+}$, whereas for large $h_0$ it approaches $1$ as $z \approx 1+\tfrac{1}{2h_0}$, $h_0 \to \infty$.

In the actual calculations we considered three points of the paramagnetic phase, $h_0=2$ ($z=1.747655$), $h_0=3$ ($z=1.33542$) and $h_0=4$ ($z=1.2112289$). In the free-fermion calculation of the gaps the lengths of the systems were $L=16,24,32,48$ and $64$, whereas with the SDRG method larger systems up to $L=512$ are treated. In all cases $10^{10}$ independent samples are investigated.

\subsection{SDRG gaps and the Fr\'echet distribution}
\label{sec:SDRG_Frechet}

\begin{figure}[h!]
\begin{center}
\hskip 1cm
\includegraphics[width=8.cm,angle=0]{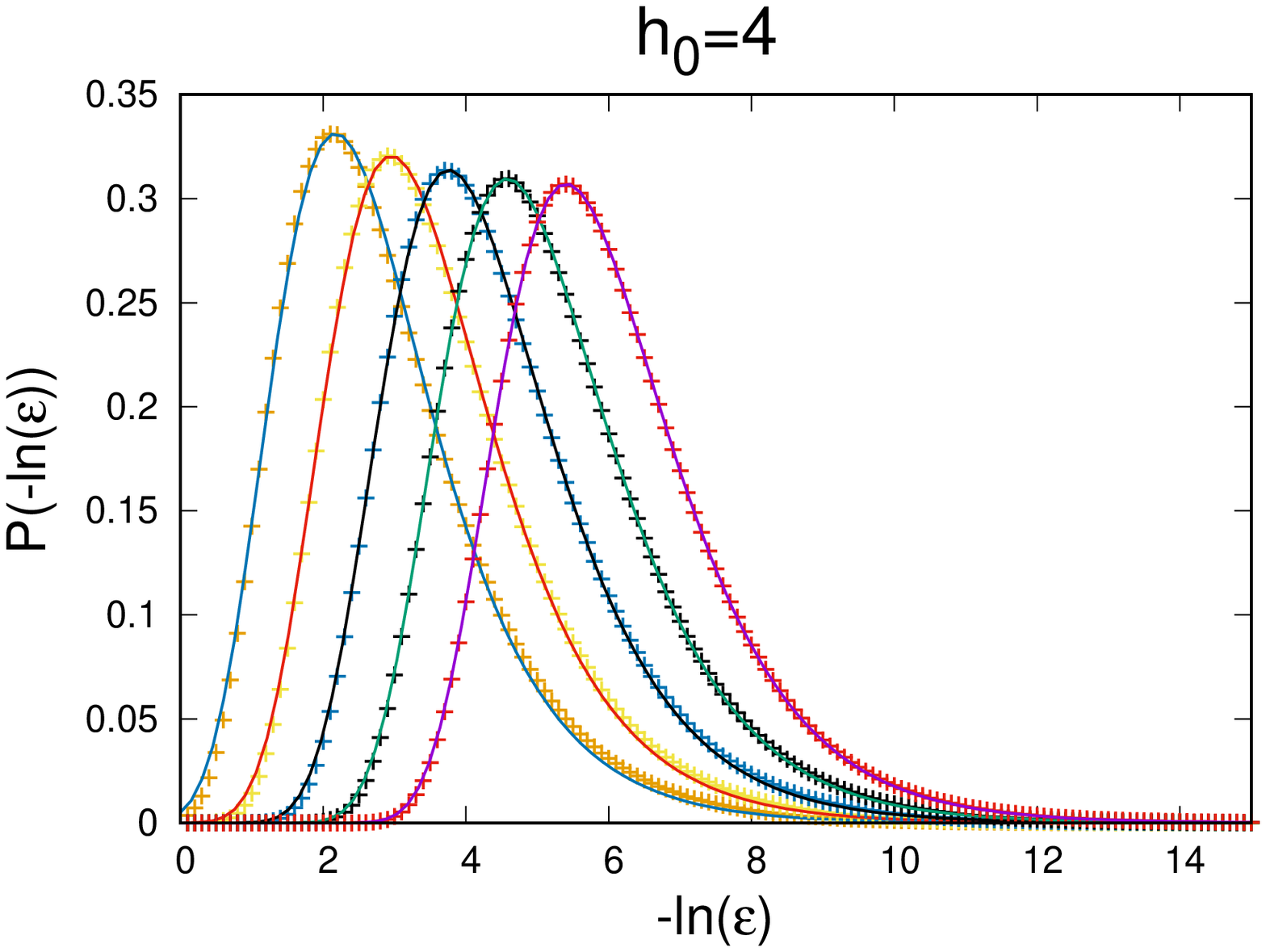}
\includegraphics[width=8.cm,angle=0]{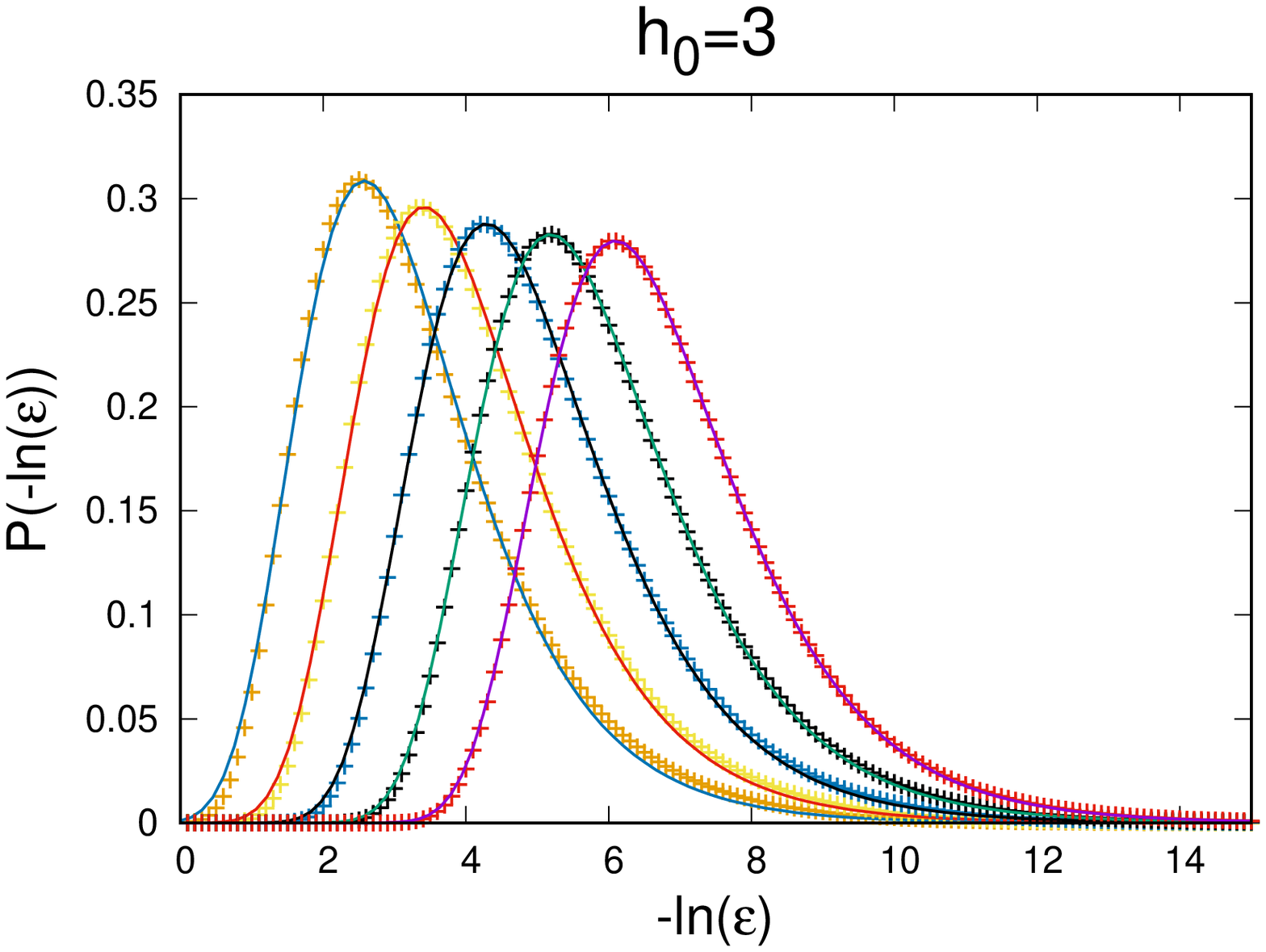}
\includegraphics[width=8.cm,angle=0]{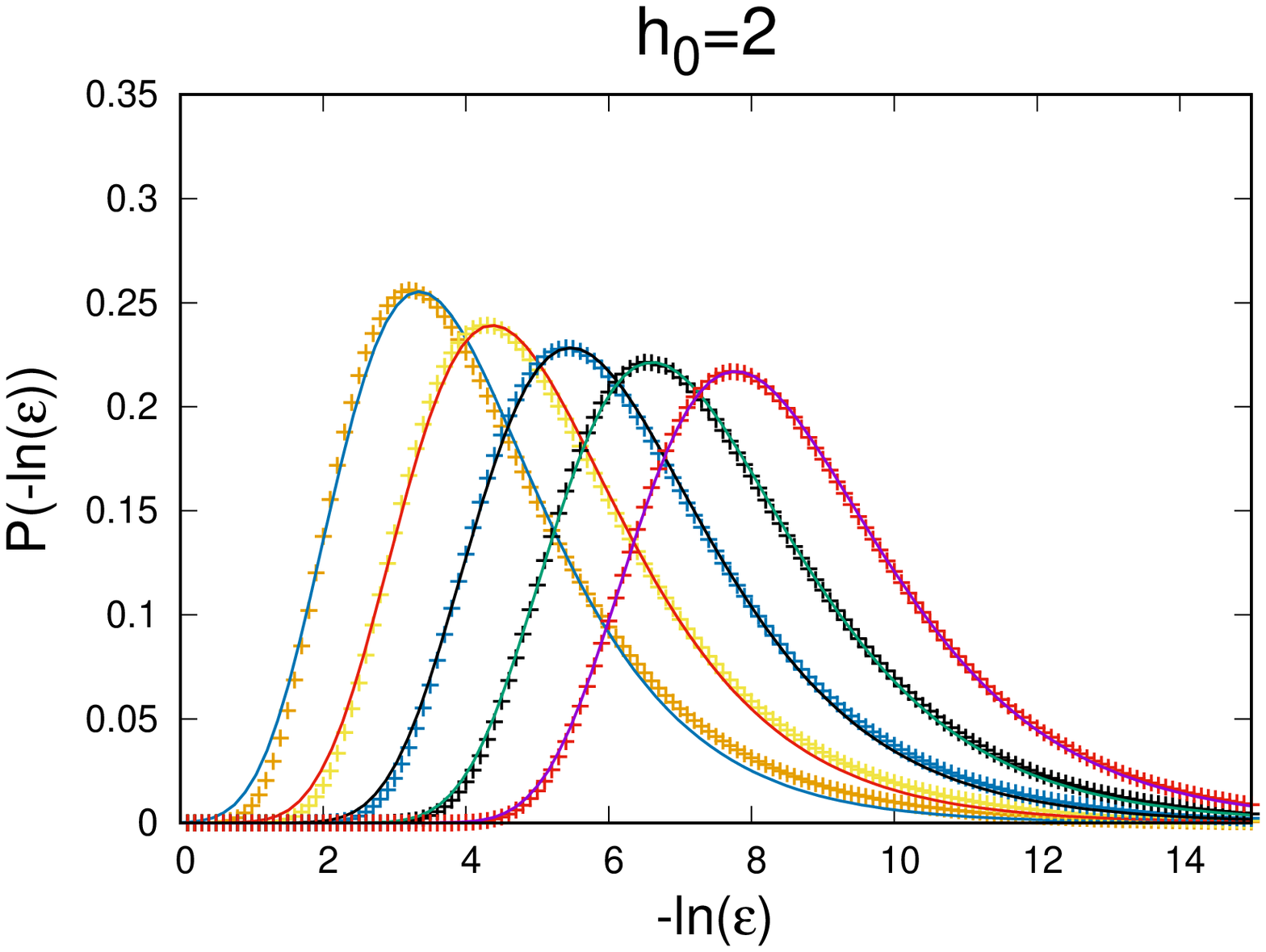}
\end{center}
\caption{\label{fig:gap_distr}(Color online) Distribution of the energy gaps calculated by the SDRG algorithm in the Griffiths phase with $h_0=4$ (upper panel), $h_0=3$ (middle panel) $h_0=2$ (lower panel) for finite chains of lengths $L=32,64,128,256$ and $512$, from left to right. The best fit of the Fr\'echet distribution is indicated by full lines.}
\end{figure}

We start by analysing the data collected by the SDRG method and presenting the distributions of the log-gaps in Fig.\ref{fig:gap_distr} for different sizes. To test their relation with the Fr\'echet distribution, we calculated in each case the best fit of the analytical function in Eq.(\ref{eq:Frechet}), having the dynamical exponent, $z_L \equiv \gamma_L$ and the position of the maximum, $x_0=x_0(L)$ as fit parameters. These are shown in Fig.\ref{fig:gap_distr}, too.

As seen in Fig.\ref{fig:gap_distr}, the Fr\'echet distribution describes the numerical data well and the difference between the numerical points and the fitting curve is decreasing for increasing values of $L$. At a given value of $h_0$ the fitted value of the dynamical exponent, $z_L$, has a small variation with the size, and approaches the exact asymptotic value, as given in Eq. (\ref{eq:z_box}). At the value $h_0=4$ this is illustrated in Table \ref{table:z(L)}, where the second column shows the fitted, finite size values of $z_L$, while the third column shows the difference from the exact value. As a first step, we used a power-law form $z-z_L \sim L^{-a}$ to fit the finite-size corrections, and estimates for the exponent $a$ are calculated through two-point fit. These are listed in the fourth column of the table and have a slow convergence, which generally indicates a strong correction to scaling term. Having this possibility in mind we have used another functional form:
\be
z-z_L \sim \frac{\ln^{\omega}L}{L}\;. 
\label{eq:log_corr}
\ee
In this case the effective exponents for $\omega$ are also calculated through two-point fit and are presented in the fifth column of the table. This type of fitting turned out to be more stable, the effective exponents seem to converge to $\omega \approx 1.33$.   

\begin{table}[ht]
\centering 
\begin{tabular}{c c c c c} 
\hline\hline 
$L$ & $z_L$ & $z-z_L$ & $a$ & $\omega$\\ 
\hline 
32 & 1.102788 & 0.108441 &  & \\ 
64 & 1.142667 & 0.068562 & 0.6614 & 1.4081 \\
128 & 1.169523 & 0.041706 & 0.7172 & 1.3724 \\
256 & 1.186572 & 0.024657 & 0.7583 & 1.3405 \\
512 & 1.196923 & 0.014306 & 0.7854  & 1.3388 \\ 
\hline 
\end{tabular}
\caption{Finite-size estimates of the dynamical exponent, $z_L$, at $h_0=4$ calculated by the SDRG algorithm from the best fit of the Fr\'echet form in Eq.(\ref{eq:Frechet}) and its difference from the asymptotically exact value $z=1.2112288988$. Exponent of a power-law fit, $a$ and that of the logarithmic correction, $\omega$. (See text.)}
\label{table:z(L)} 
\end{table}

We have repeated the same analysis at the other two points of the Griffiths phase. In both cases the log-correction form is found to provide the better fit, with the correction exponents $\omega \approx 1.83$ at $h_0=2$ and $\omega \approx 1.50$ at $h_0=3$.

Most importantly, at a given value of $h_0$ in Fig.\ref{fig:gap_distr} the distributions are shifted with increasing size, and according to Eq.(\ref{L_z}) the position of the maximum is expected to follow the rule: $x_0(L) \approx const.+ z_L \ln L$. Comparing the position of the maximum of the distribution at two sizes one can obtain estimates for the dynamical exponent as: $z(L,2L)=(x_0(2L)-x_0(L))/\ln 2$. We have checked that generally $z_L<z(L,2L)<z_{2L}$ and these estimates approach the asymptotic exact value with the same type of corrections as noticed for the case of $z_L$ in the previous paragraph.

We note that analysis of the data obtained by free-fermion calculation of the gap gives similar results, but due to smaller values of $L$ the asymptotic region of the effective exponents, $z_L$ is more remote.

\subsection{Finite-size corrections to the gap distributions}
\label{sec:finite_size}
\begin{figure*}[t]
\centering
  \begin{tabular}{@{}ccc@{}}
     \includegraphics[height=.235\linewidth]{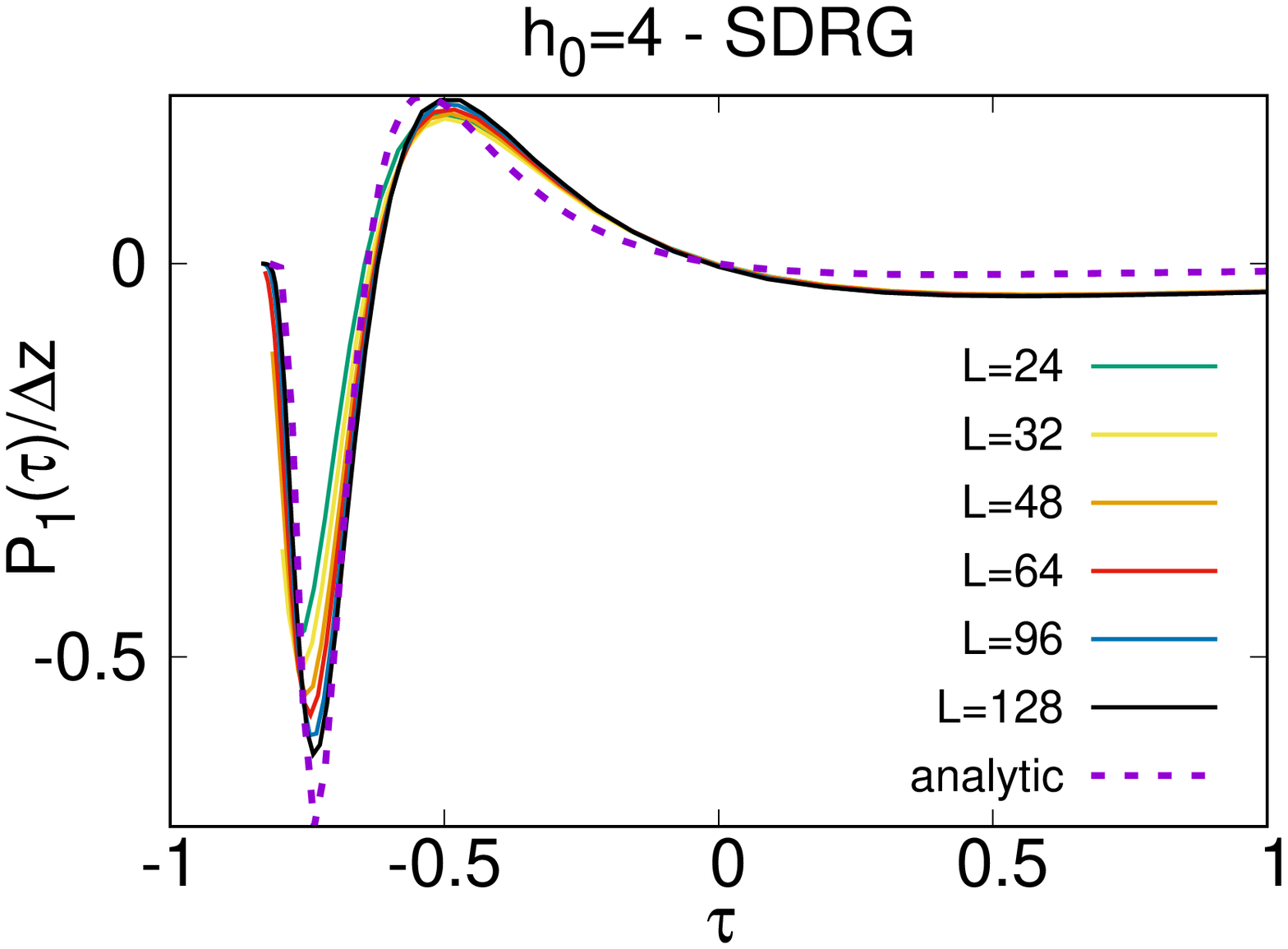} &
     \includegraphics[height=.235\linewidth]{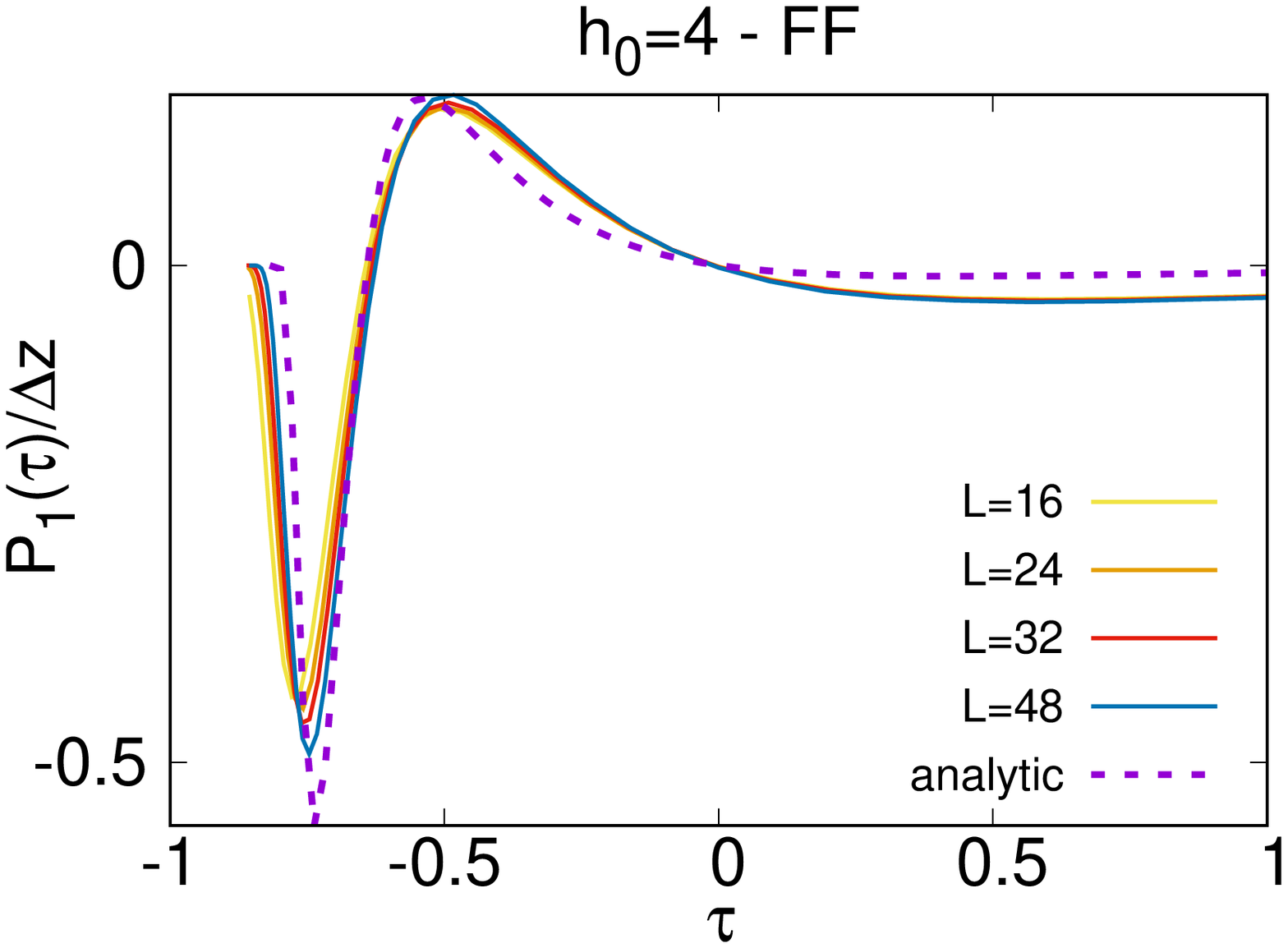} &
    \includegraphics[height=.235\linewidth]{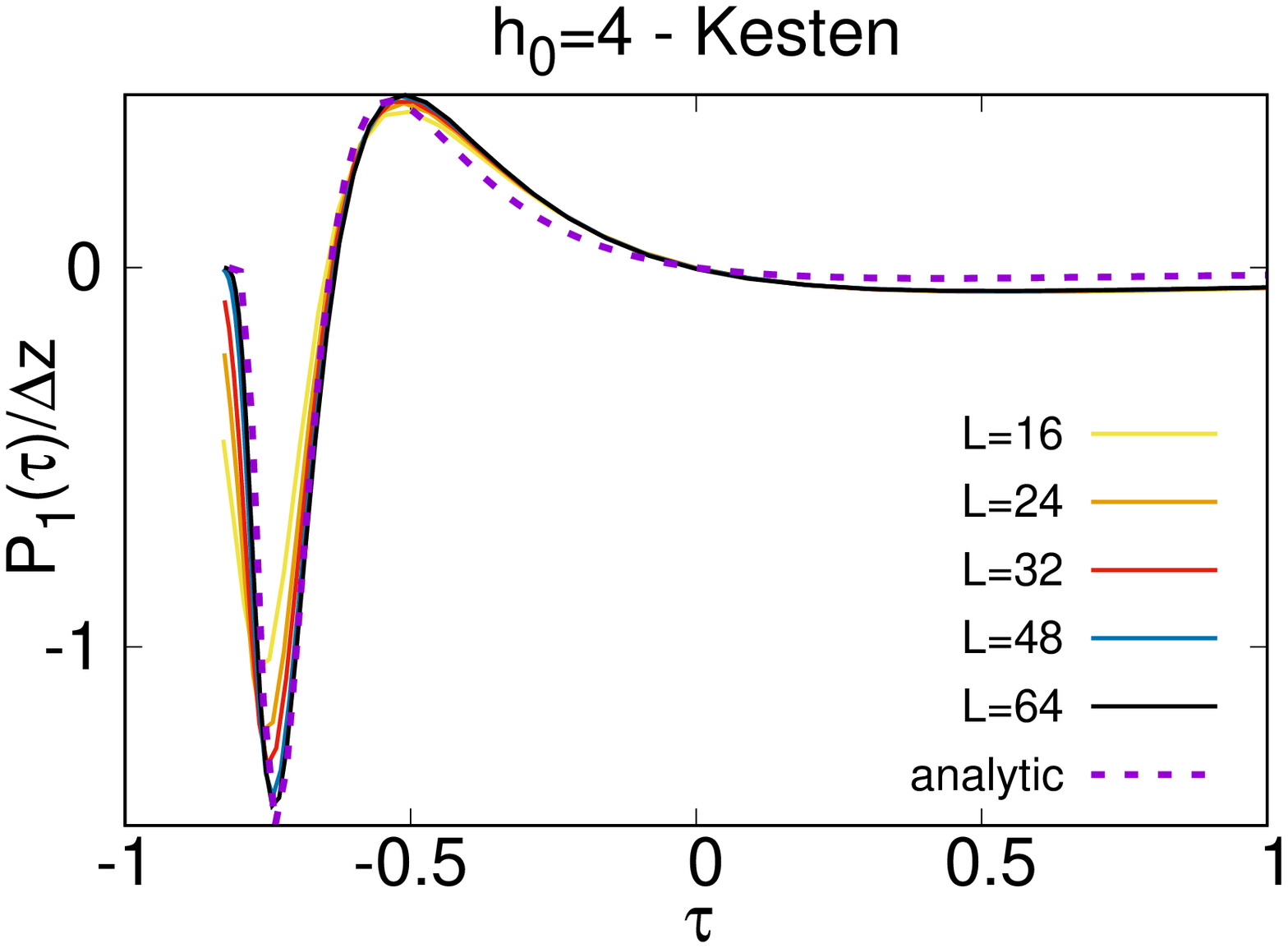}\\
    \includegraphics[height=.235\linewidth]{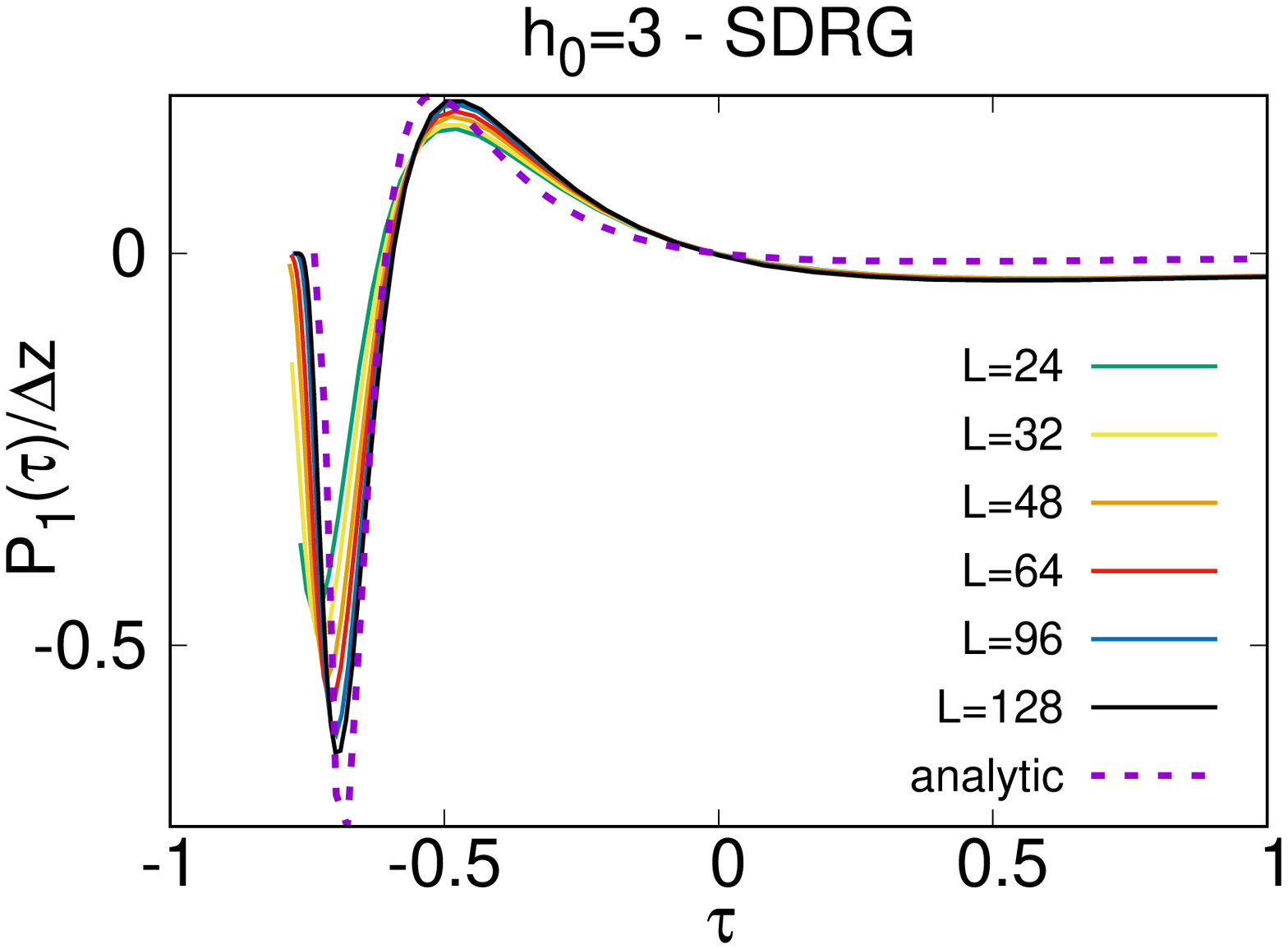} &
    \includegraphics[height=.235\linewidth]{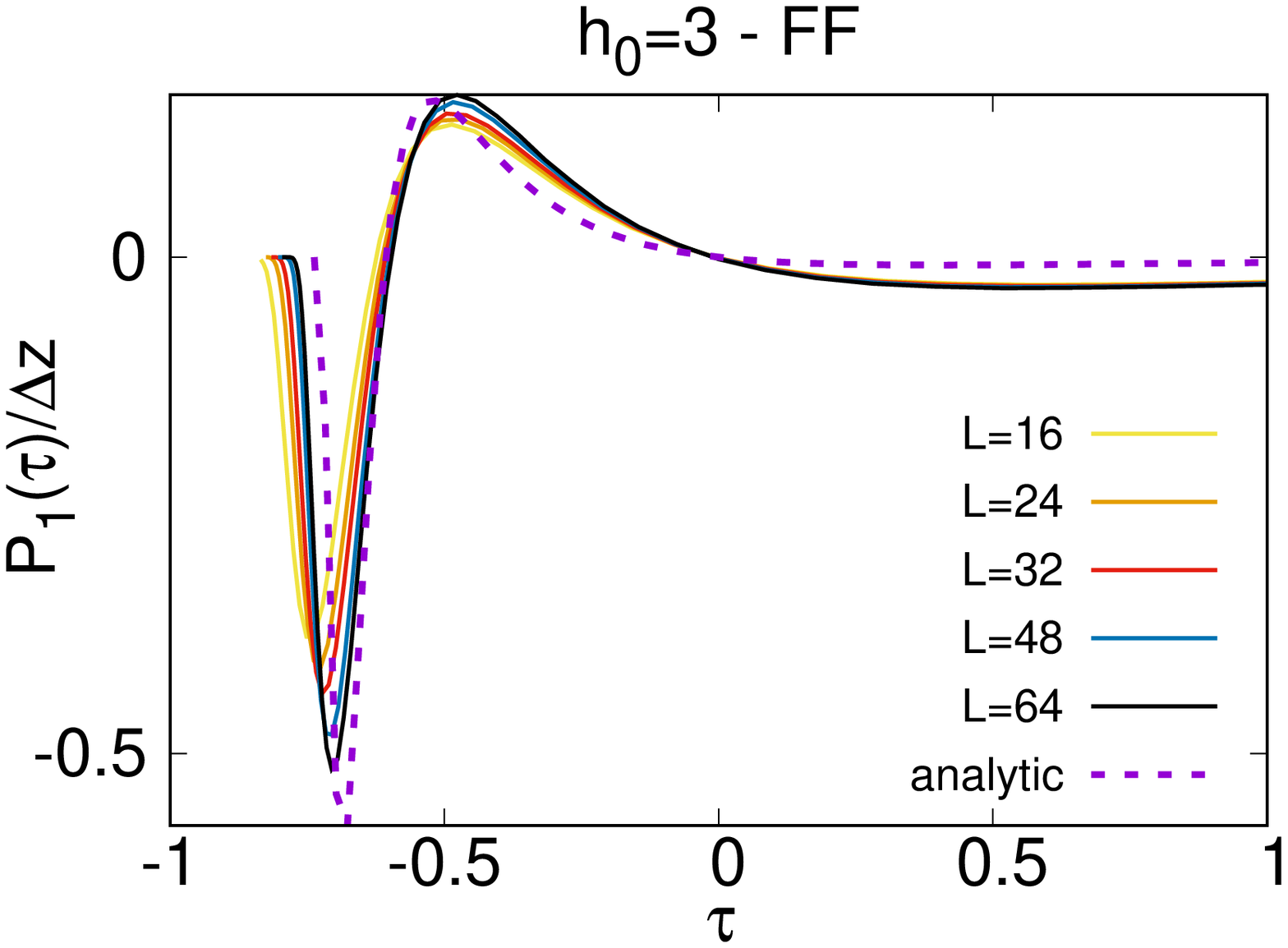} &
    \includegraphics[height=.235\linewidth]{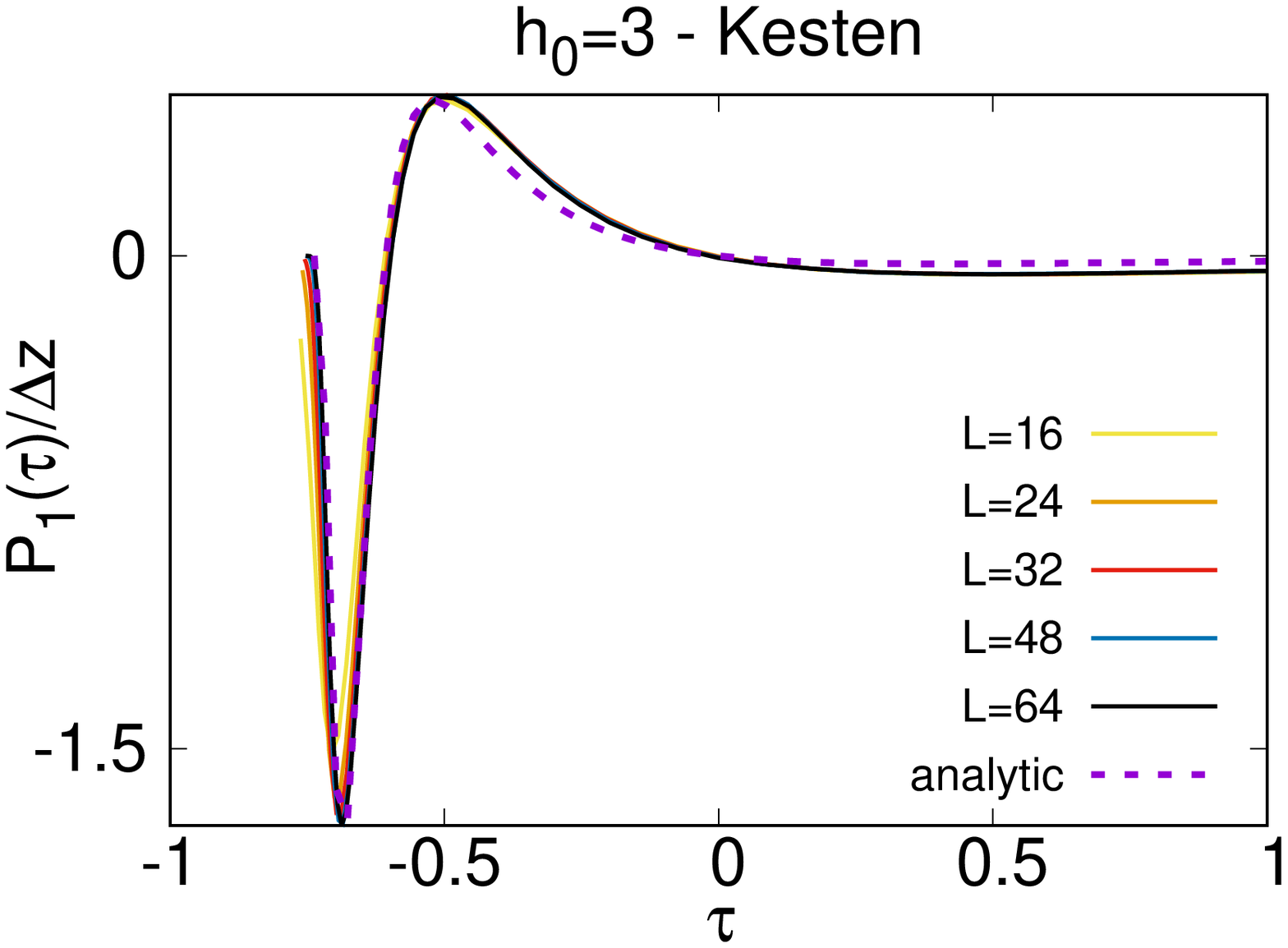} \\
    \includegraphics[height=.235\linewidth]{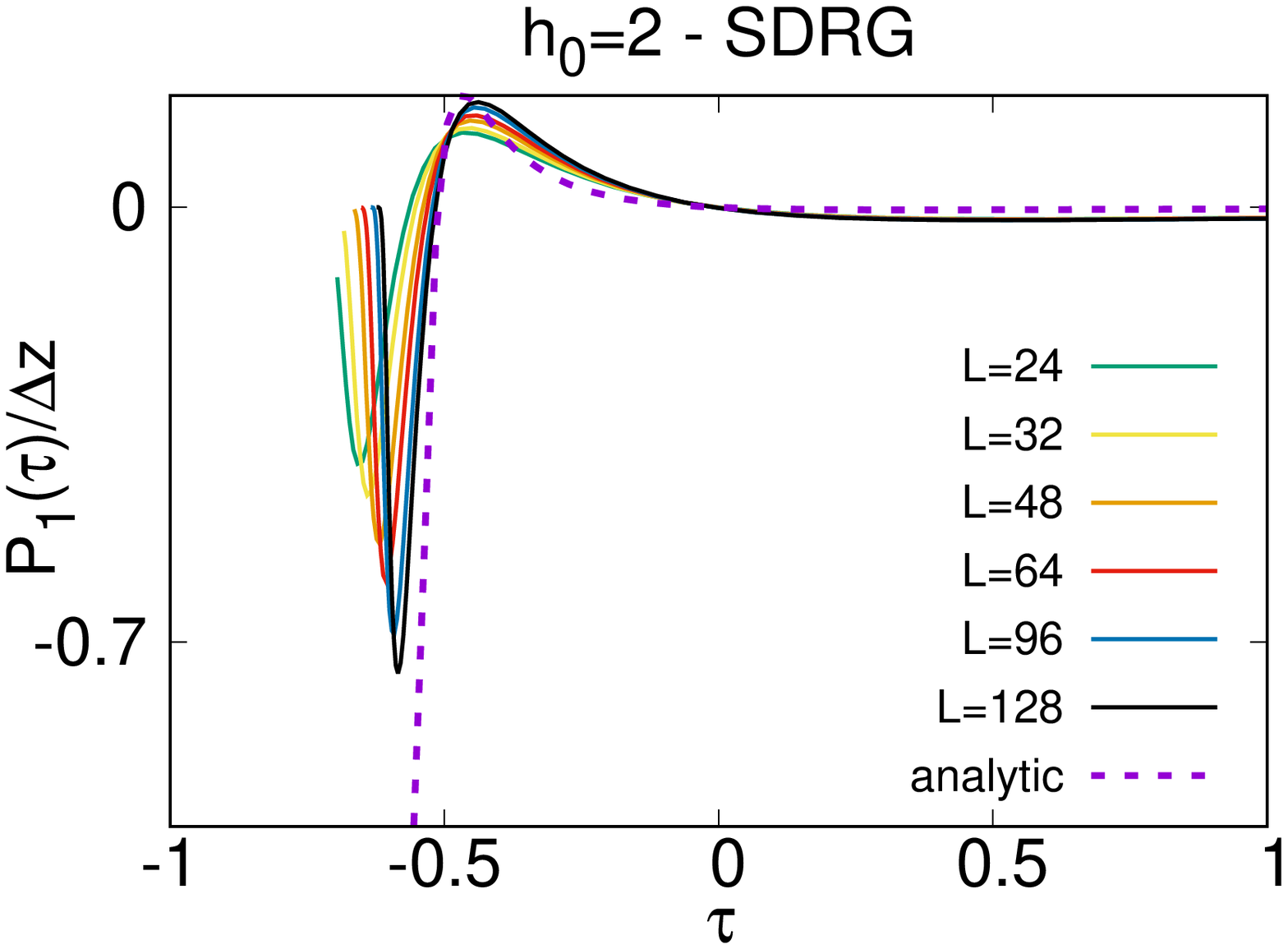} &        
    \includegraphics[height=.235\linewidth]{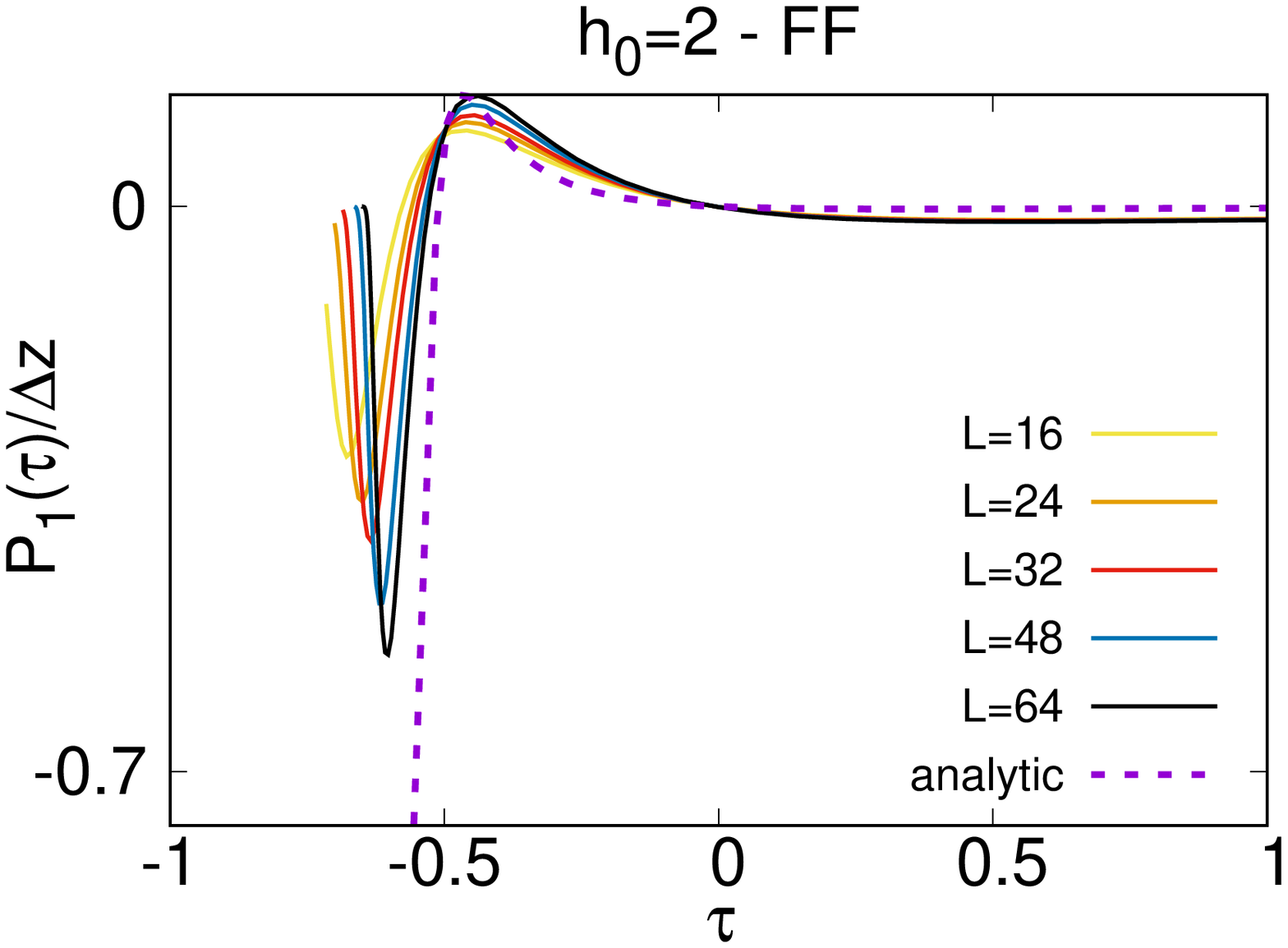} &
    \includegraphics[height=.235\linewidth]{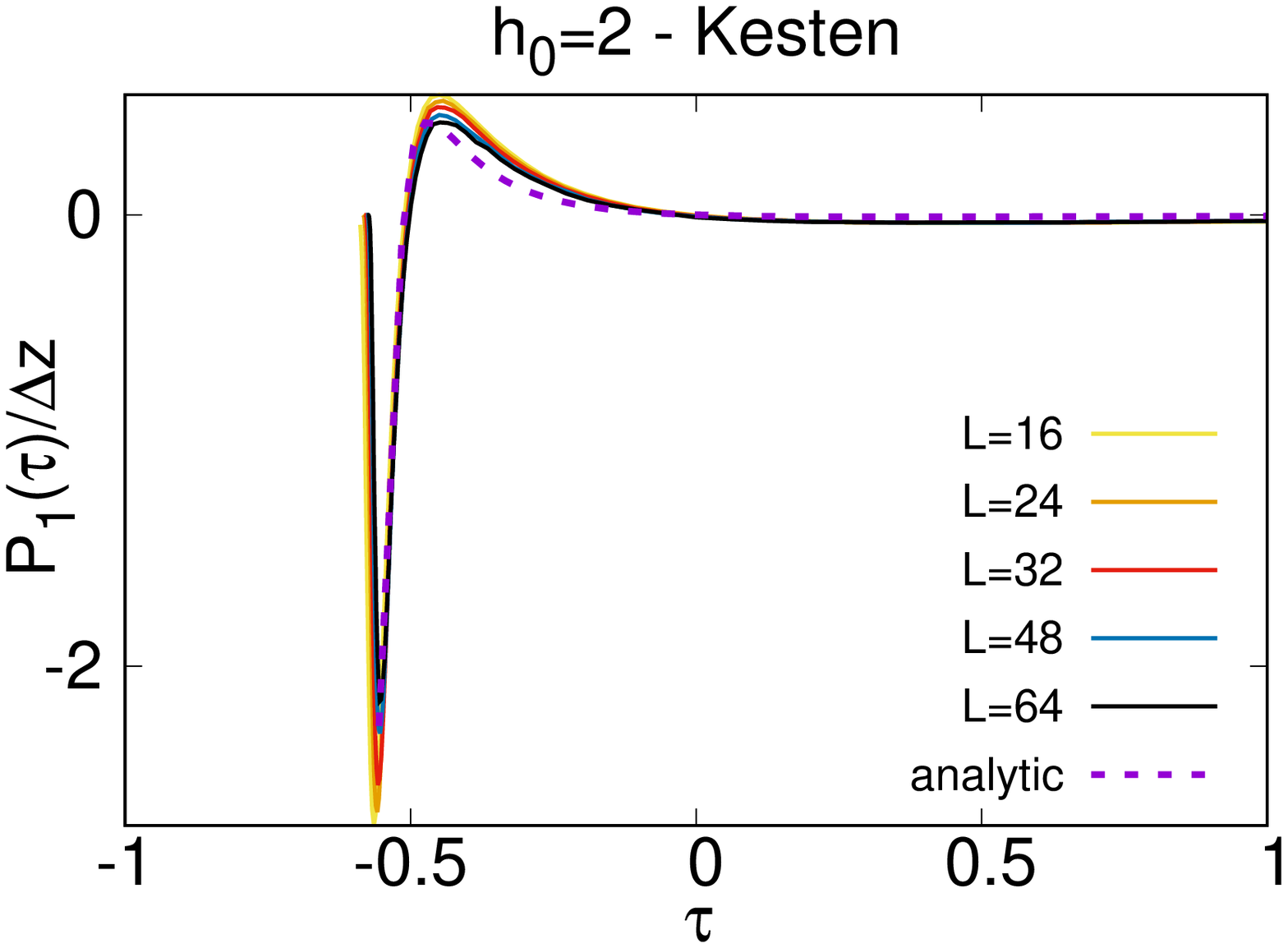} \\
  \end{tabular}
  \caption{(Color online) Finite-size corrections to the Fr\'echet distribution for the inverse gaps, $\tau=1/\varepsilon$ in the RTIM at different sizes $L$ calculated through SDRG iterations (left panels), and by the free-fermion method (middle panels) with the first standardisation in Eq.(\ref{1st_standardisation}) and compared with the analytical results with $\gamma=z(h_0)$ and $\gamma'=-1$ (dashed lines). In the right panels the maximum of uncorrelated Kesten random variables are shown, having the same scaling exponents, see text. The scaling exponents $z(h_0)$ correspond to $h_0=4$ (first row), $h_0=3$ (second row) and $h_0=2$ (third row), see in Eq.(\ref{eq:z_box}).
  \label{fig:RTIM_1st}}
\end{figure*}
We considered the gaps calculated by SDRG iteration for $L=24,32,48,64,96$ and $128$ and for comparison we calculated those by free-fermion techniques as well, for shorter chains with $L=16,24,32,48$ and $64$, except for $h_0=4$, where the corrections are the smallest and we went up to $L=48$.
We analysed finite-size scaling of distributions in two different ways. First, we used the \emph{first standardisation convention} in Eq.(\ref{1st_standardisation}) and calculated the difference with the Fr\'echet extreme 
distribution, with an effective $z_L$, calculated from the relations below Eq.(\ref{eq:Frechet}). This difference is then rescaled by a factor $\Delta z=z-z_L$, and the results for $h_0=2,3$ and $4$ are drawn in Fig.\ref{fig:RTIM_1st}. In these figures the analytical results calculated for iid random numbers in Eq.(\ref{eq:psi_v}) with $\gamma=z$ and $\gamma'=-1$ are also presented. Here, the correction to scaling exponent, $\gamma'=-1$ corresponds to the expected scaling form in Eq.(\ref{eq:log_corr}), which is obtained through numerical analysis of the data in Table \ref{table:z(L)}. According to Fig.\ref{fig:RTIM_1st}, we can draw the following conclusions. i) The scaled finite-size difference of the distribution function seems to approach a limiting curve for large $L$, which depends on $z(h_0)$. ii) At a given $z(h_0)$ the limiting curves are similar (if not identical) for the SDRG and the free-fermion data. iii) The convergence to this limit curve is slow, much slower than that of the iid random numbers, see later in Sec.\ref{sec:num_iid} and the figures in the third column of Fig.\ref{fig:RTIM_1st}. This slow convergence is probably related to the logarithmic correction to the scaling of the dynamical exponent. iv) Finally, the expected limit curve of the numerical distribution differs from that of the analytical result calculated for iid random numbers (having identical parameter, $z$). Even though the overall shapes of the curves are similar, there are noticeable differences. In particular, the low-energy part of the numerical curves are stronger represented in the numerical curves, which can be interpreted as the reduction of the value of the gap due to small, but relevant correlations between the rare regions.

\begin{figure*}[t]
\centering
  \begin{tabular}{@{}ccc@{}}
    \includegraphics[height=.235\linewidth]{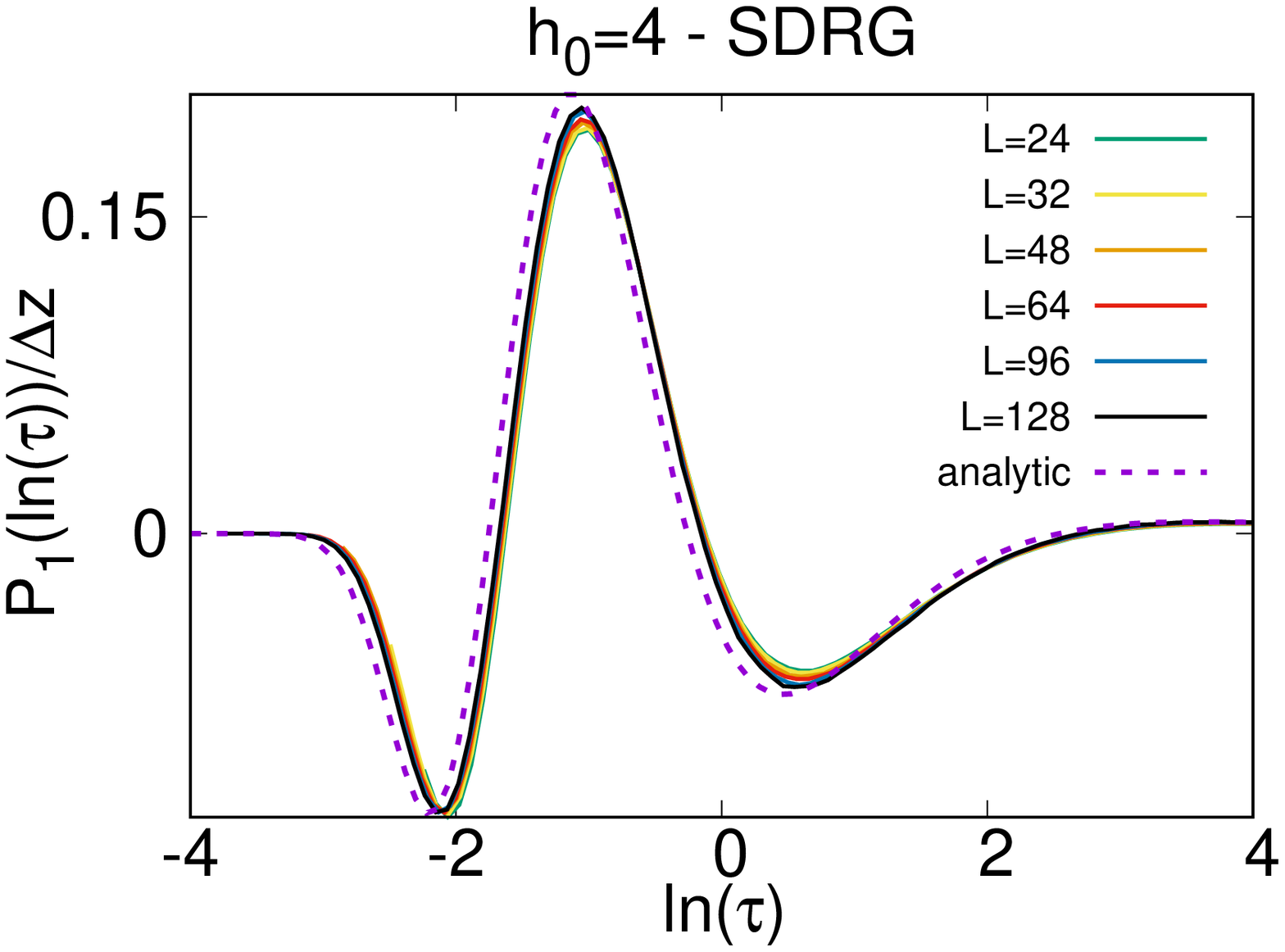} &
    \includegraphics[height=.235\linewidth]{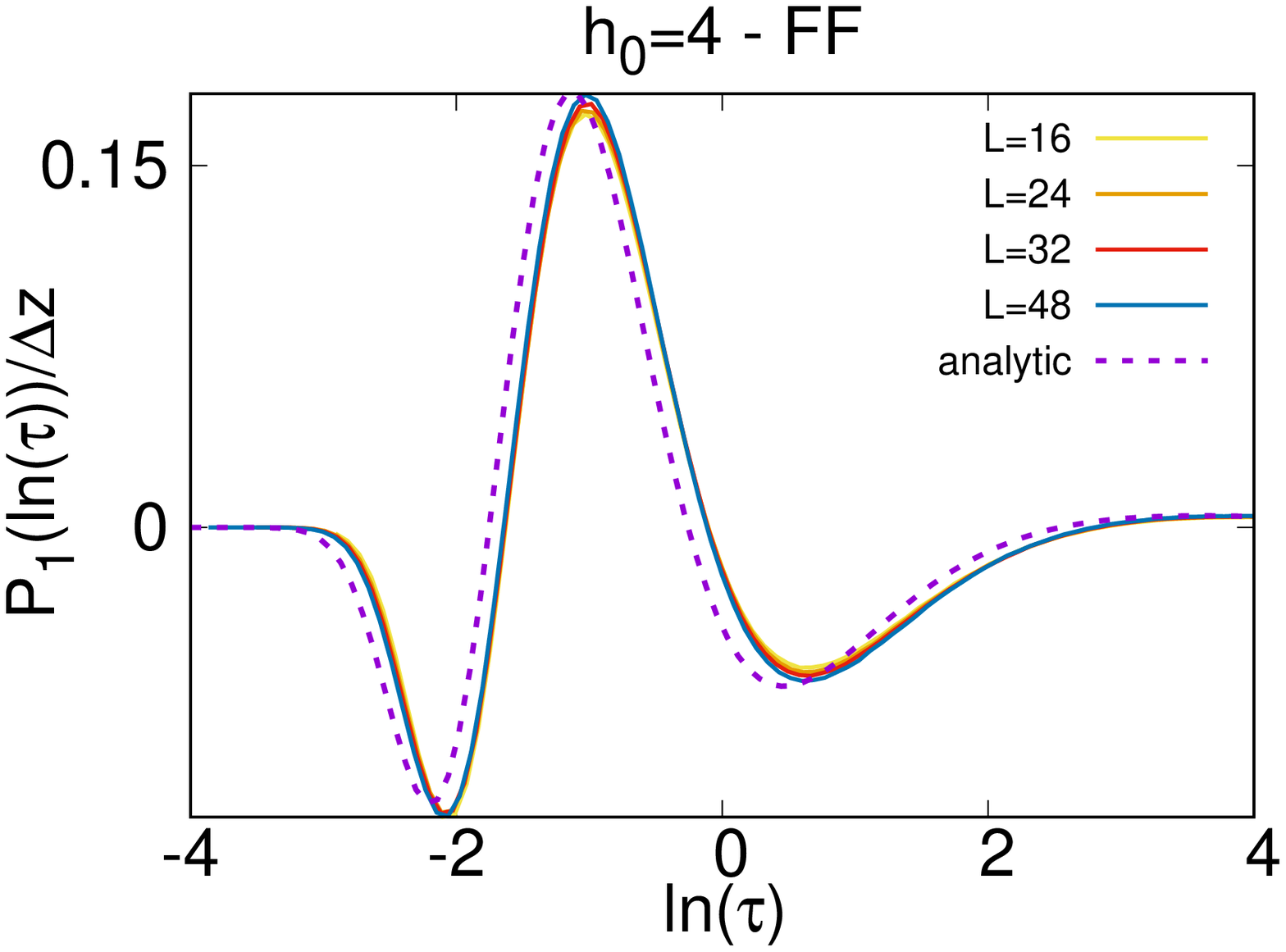} &
    \includegraphics[height=.235\linewidth]{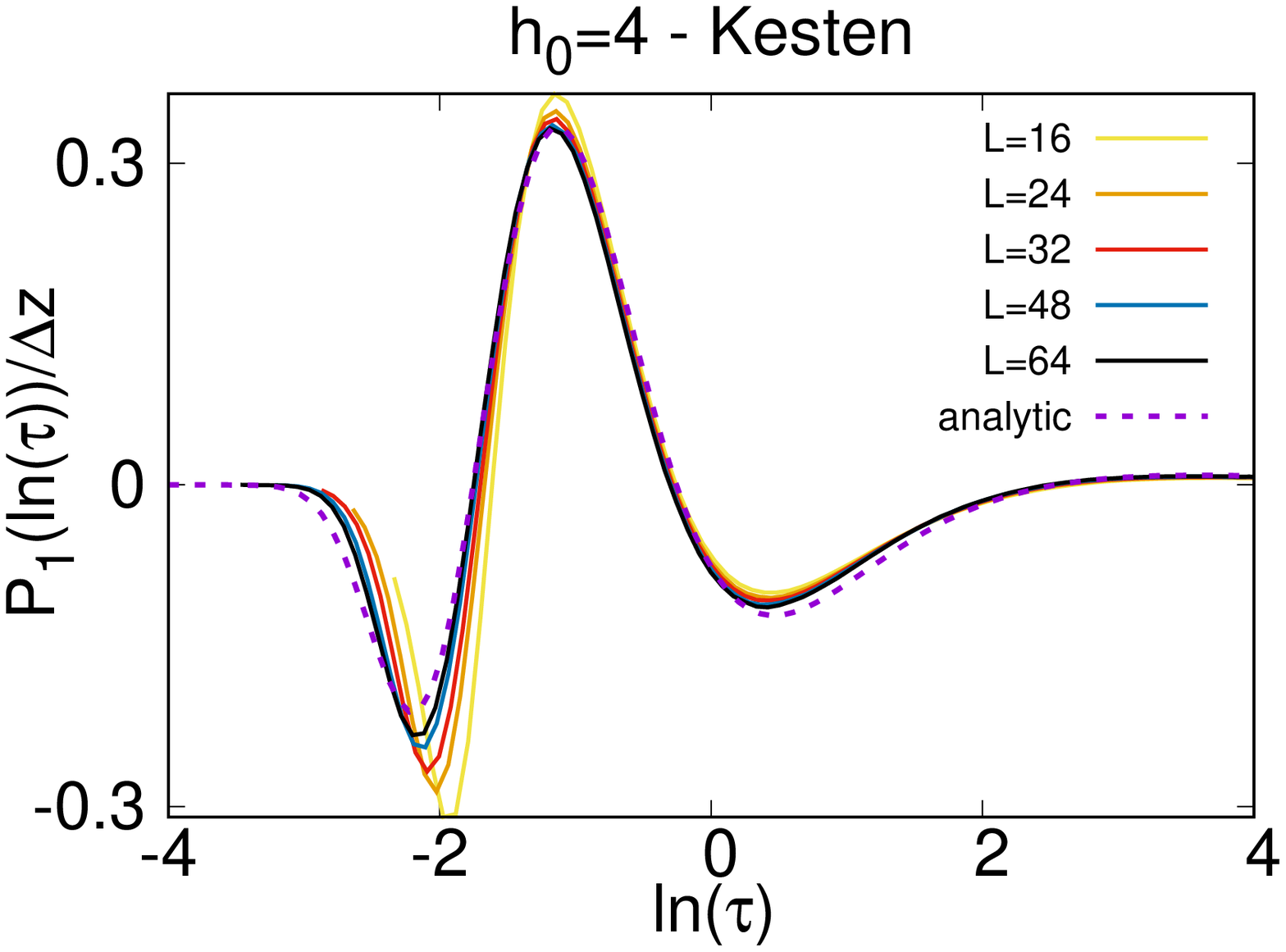} \\
    \includegraphics[height=.235\linewidth]{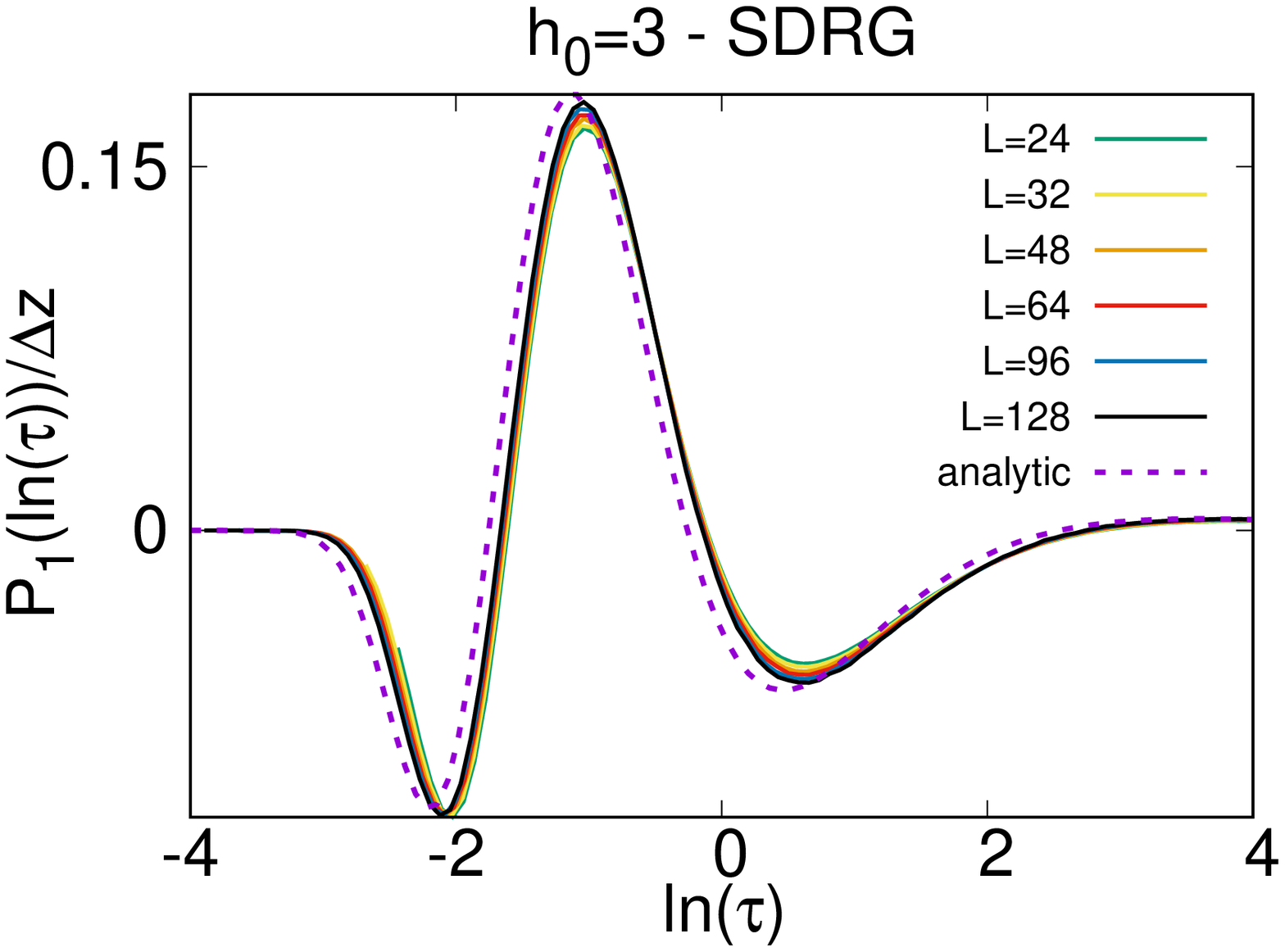} &
    \includegraphics[height=.235\linewidth]{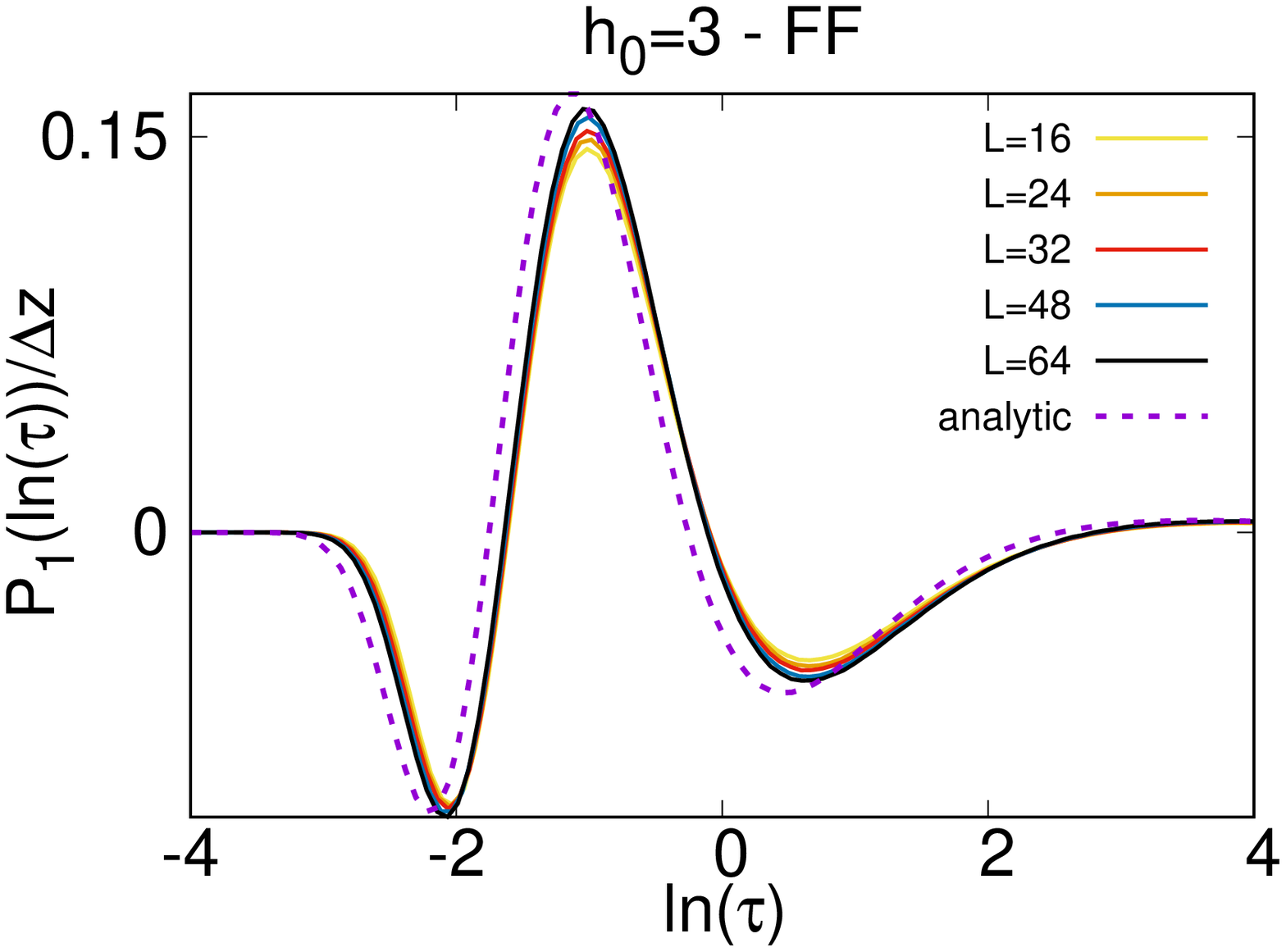} &
    \includegraphics[height=.235\linewidth]{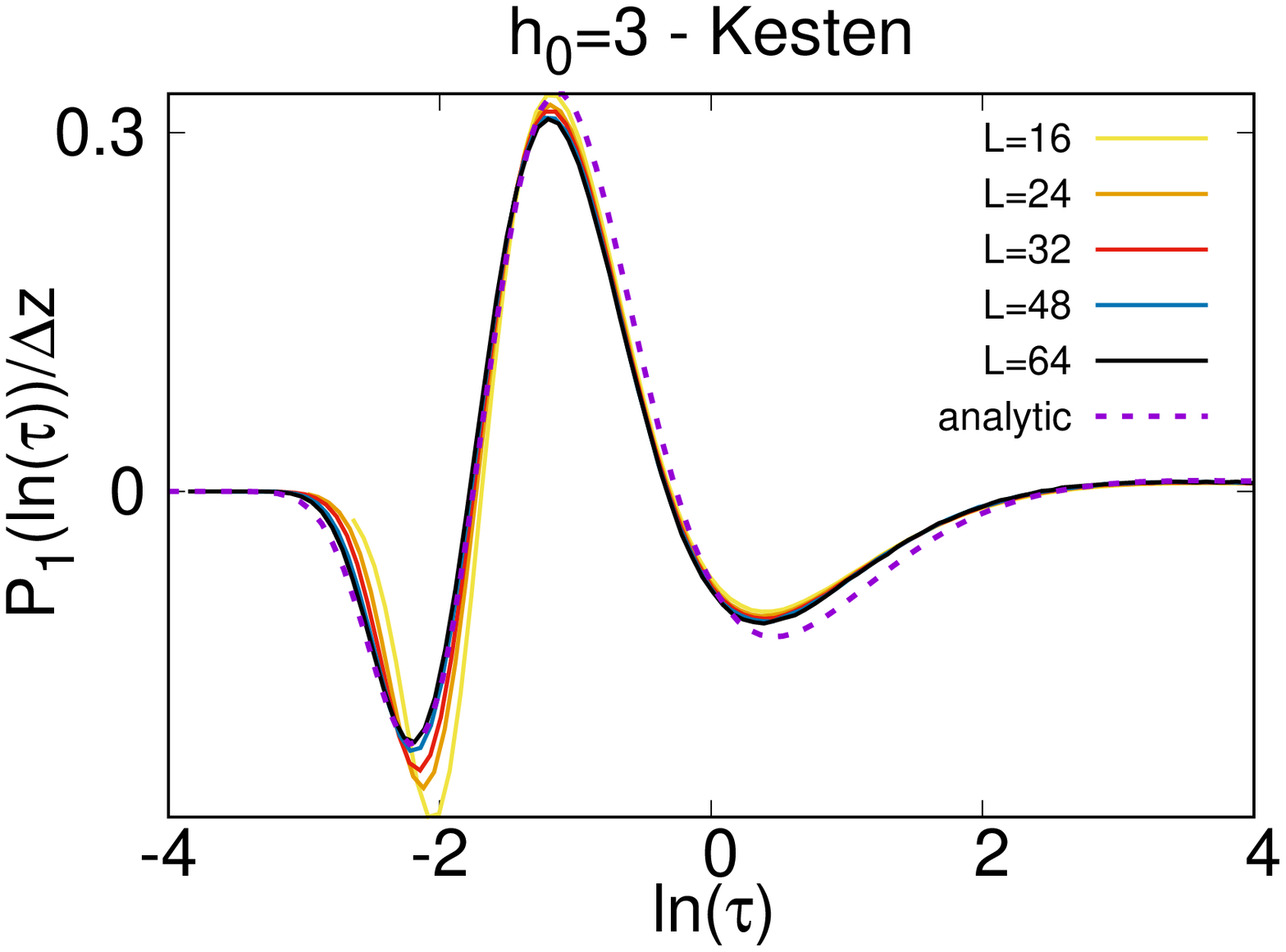} \\
     \includegraphics[height=.235\linewidth]{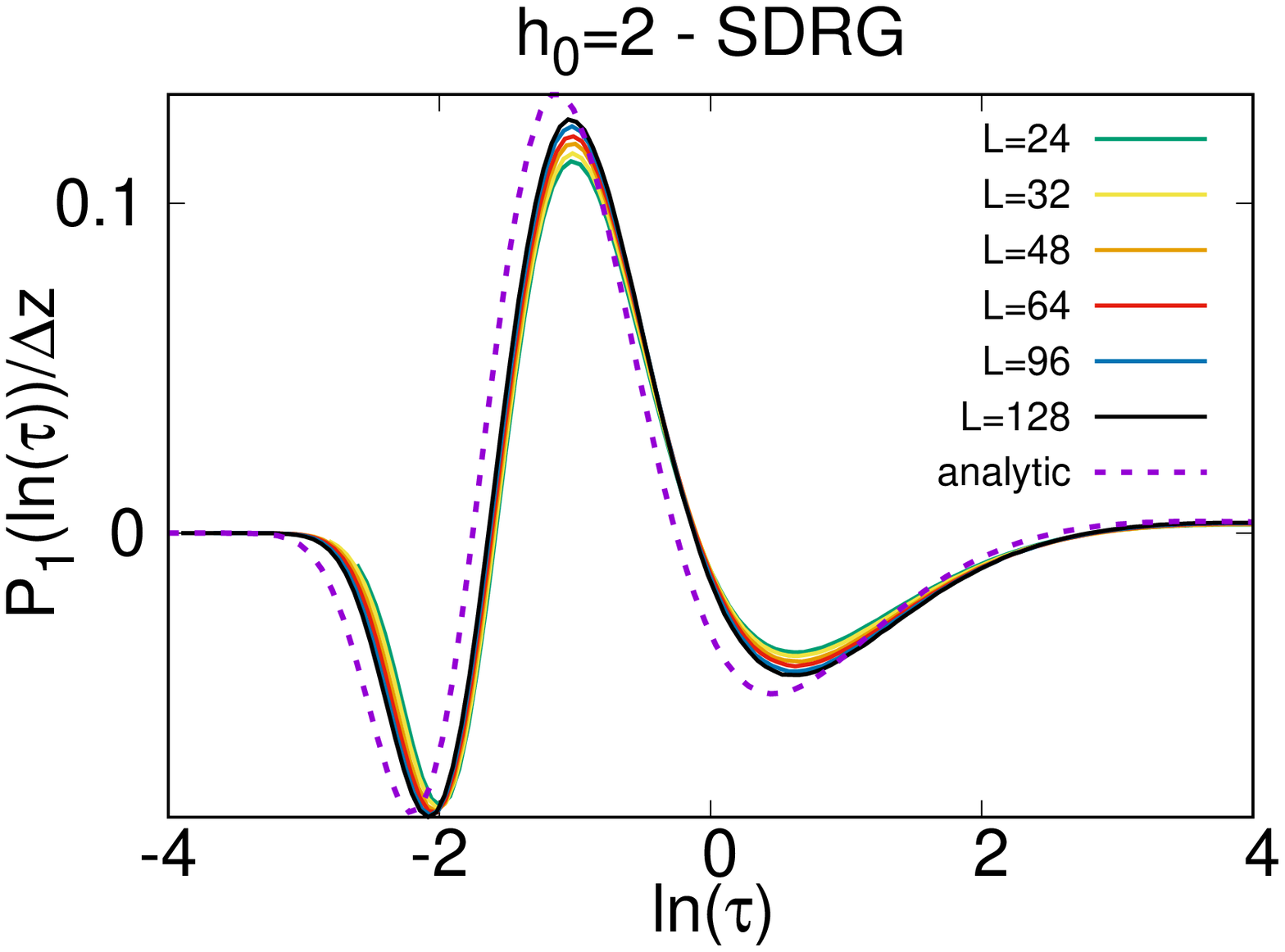} &
    \includegraphics[height=.235\linewidth]{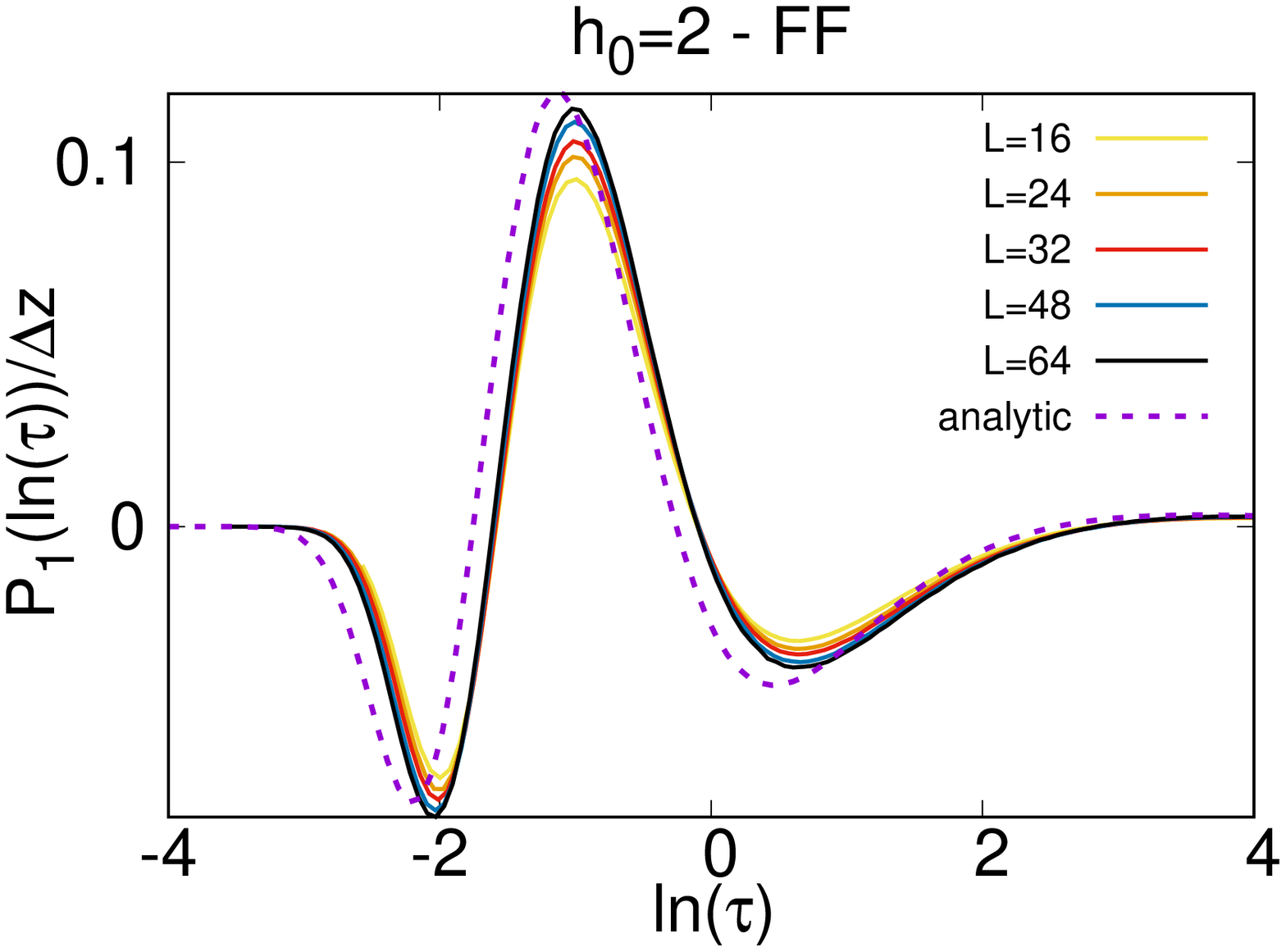} &
    \includegraphics[height=.235\linewidth]{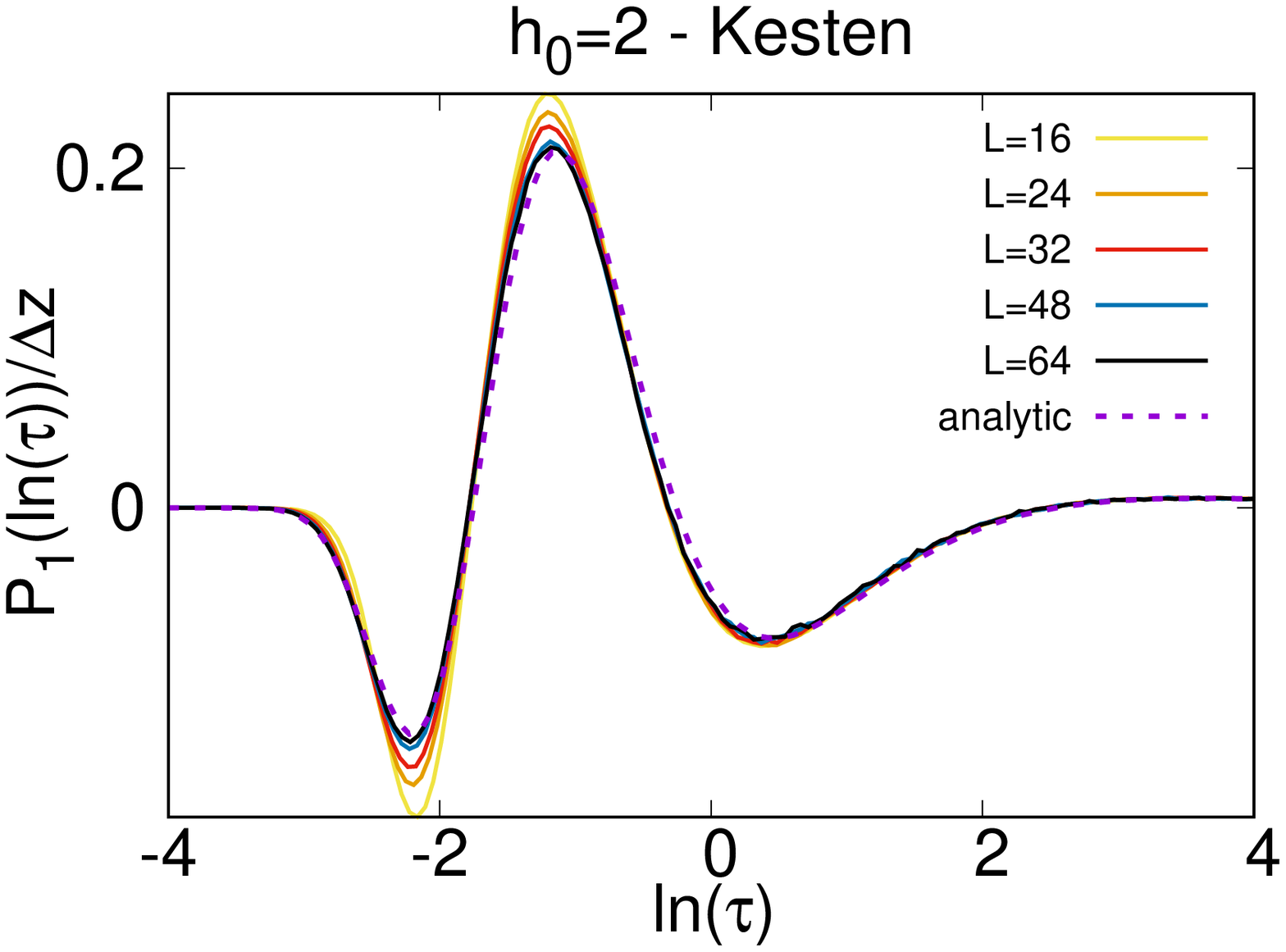}
  \end{tabular}
  \caption{(Color online) The same as in Fig.\ref{fig:RTIM_1st} for the log-variable and using the second standardisation condition in Eq.(\ref{2nd_standardisation}).
  \label{fig:RTIM_2nd}}
\end{figure*}

We have repeated the analysis with the distribution of the log-gaps and using the second standardisation condition in Eq.(\ref{2nd_standardisation}). The results about the finite-size
 corrections of the distributions are shown in Fig.\ref{fig:RTIM_2nd} together with the analytical curves for iid random 
 numbers. In Fig.\ref{fig:RTIM_2nd} the distributions of the log-gaps have a faster finite-size convergence, than those of the gaps in Fig.\ref{fig:RTIM_1st}. The curves in Fig.\ref{fig:RTIM_2nd} are almost indistinguishable for larger sizes. In addition, we notice a difference between the shape of the numerical curves and the analytical results, the former being somewhat shifted to the right around zero. This indicates a reduction of the value of the gap due to small, but relevant correlations between the rare regions.

\subsection{Numerical test for uncorrelated Kesten variables}
\label{sec:num_iid}

In order to test the possible convergence of the finite-size corrections for uncorrelated variables we have repeated 
the analysis in the previous subsection for a parent distribution generated by Kesten random numbers. Here, we remark that Kesten-type random variables are defined as\cite{kesten73}:
\be
u_m=1+\sum_{i=1}^m \prod_{j=1}^i s_j+\dots\;,
\label{kesten}
\ee
where the $s_j$-s are iid random numbers. It is known, that in the limit of $m \to \infty$ there is a limit distribution, $\rho_{\infty}(u)$, provided $[\ln s]_{\rm av}<0$. This limit distribution has a power-law tail
\be
\rho_{\infty}(u) \sim u^{-(1+\alpha)};\quad u\gg 1\;,
\ee
where the exponent is the positive root $\alpha>0$ of the equation
\be
\left[s^{\alpha}\right]_{\rm av}=1\;.
\ee
Note that the relation for $\alpha$ is analogous to the equation for the dynamical exponent, $z$ of the RTIM, see in Eq.(\ref{eq:z_eq}), and so we have the relation $\alpha=1/z$. To have a direct relation with the RTIM calculations we set $s_j=J_j/h_j$ where $J_j$ and $h_j$ are taken from the distributions in Eq.(\ref{eq:J_distrib}).
In the numerical calculation, we have $m=64$, and considered the maximum of a set of $L=16,24,32,48$ and $64$ Kesten numbers, denoted by $\tau$, so that we have $m \ge L$ in each case. We have checked that at such value of $m$ the truncation of the series in Eq.(\ref{kesten}) has negligible error.

We have analysed the distribution of the maximum values at the three points $h_0=2$, $h_0=3$ and $h_0=4$ as done previously for the gaps of the RTIM. Using the first standardisation condition in Eq.(\ref{1st_standardisation}) the finite-size corrections are shown in the third column of Fig.\ref{fig:RTIM_1st}. Here, the analytical results are obtained with the exponents: $\gamma=z(h_0)$ and $\gamma'=-1$, the latter follows from the analytical results in Ref.[\onlinecite{clnp}]. It is seen in Fig.\ref{fig:RTIM_1st} that there is an overall good agreement between the numerical and analytical results. There is some size-dependence of the corrections, which changes sign between $h_0=2$ and $h_0=3$. In comparison to the distribution of the gaps in the RTIM, the agreement with the analytical results is much better and the finite-size corrections are smaller.

We have repeated the analyses of the data by considering log-variables, as given in Eqs.(\ref{eq:M_Fr_x}) and (\ref{eq:Frechet}) and using the second standardisation condition in Eq.(\ref{2nd_standardisation}). Results of the numerical analysis in this case are given in the third column of Fig.\ref{fig:RTIM_2nd}. As seen in this figure, there are some finite-size dependences of the scaled numerical data, but the expected asymptotic curves agree well with the analytical results. We note that practically no finite-size dependence of the correction term is observed for a pure power parent distribution in Fig.\ref{fig:pure_power} in the Appendix.

\section{Discussion}
\label{sec:disc}

In this paper we have considered a paradigmatic model of random quantum magnets, the random transverse Ising model in 1D and studied the distribution of low-energy excitations in the paramagnetic Griffiths phase, with extensive numerical methods. We have considered a large set of random samples ($10^{10}$) and the calculation is performed by the approximate, but asymptotically correct SDRG method (up to $L=512$) and for comparison we also used the free-fermion method for shorter chains (up to a size $L=64$). Analysing the distribution of the gaps we have demonstrated with high precision that -- in agreement with previous expectations -- the limit distribution in the thermodynamic limit is in the Fr\'echet form. The Fr\'echet distribution depends on the value of the dynamical exponent, $z$ and we have shown that a powerful method of calculation is through fitting a Fr\'echet curve to the numerical gap distributions. In this way, effective, finite-size estimates are obtained for the dynamical exponent $z_L$, which then are extrapolated to $L \to \infty$. According to the numerical data this convergence is in the form: $z-z_L \sim \ln^{\omega} L/L$, where the exponent of the logarithm, $\omega$, depends on the distance from the critical point.

More interestingly, we have systematically studied the finite-size corrections to the limit law and showed that the difference between the numerical distribution and the asymptotic Fr\'echet form scales with $z-z_L$, and this scaled difference, $P_1(\varepsilon;z)$ is a unique function of the value of the gap, $\varepsilon$. We have performed this type of analysis for the gap, using the first standardisation condition in Eq.(\ref{1st_standardisation}), as well as for the log-gap, when the second standardisation condition in Eq.(\ref{2nd_standardisation}) was used. In both cases the asymptotic form of the scaled finite-size correction function are found similar (if not identical) for the data with the free-fermion calculation and that from the SDRG iteration. We can thus conclude that the SDRG method provides not only the correct asymptotic form of the gap distribution function, but the correct finite-size correction function as well.

The measured finite-size correction functions are also compared with the analytical results of iid random numbers, having the same decay exponent ($\gamma=z$ for the gaps and $\gamma=0$ for the log-gaps) and correction to scaling exponent, $\gamma'=-1$. The two curves are found to have similar shape, but there are also differences in the asymptotic forms. We have checked that the observed differences are larger than the statistical error of the calculation. For this purpose, we have analysed with identical methods the same set of uncorrelated Kesten random variables having the same characteristic exponent, $z$. For these iid random numbers the asymptotic finite-size corrections are found to be well described by the analytical results.

The observed difference in the finite-size corrections between the numerical curves and the analytical iid results indicates that the weak correlations between low-energy excitations in the RTIM are not completely irrelevant. This is probably related to the fact that the linear extension of rare regions, $n$, with extreme fluctuations of the strong couplings, scales as $n \sim \ln L$, see the reasoning above Eq.(\ref{M_tau}). Since $n$ is not limited, the finite-size dynamical exponents contain a logarithmic multiplicative factor as given in Eq.(\ref{eq:log_corr}).

Our investigations have concluded on the RTIM in 1D, but the observed results are possibly valid for other random quantum systems as well with localised excitations. For example, we can mention the RTIM in higher dimensions\cite{2d,2dki,ddRG}, the random quantum Potts\cite{senthil} and Ashkin-Teller models\cite{carlon-clock-at} and generally random quantum magnets with short range interactions and with discrete symmetry. Similar conclusions apply to some non-equilibrium processes, such as to the random asymmetric exclusion process\cite{igloipartial} and the random contact process\cite{igloicontact} as well.

\begin{acknowledgments}
This work was supported by the National Research Fund under Grants No. K128989, No. K115959 and No. KKP-126749. F.I. is indebted to Katalin Ozog\'any for sending her PhD thesis Ref.[\onlinecite{ozogany}] and for valuable correspondence.
\end{acknowledgments}

\section*{Appendix - Extreme statistics of uncorrelated variables}
\label{sec:extreme_statistics}

Let us have a process of iid random numbers $u_1,u_2,\dots,u_N$ and consider the maximum value $u={\rm max}(u_1,u_2,\dots,u_N)$. Each random variable is distributed by the same, so called \emph{parent} distribution function $\rho(u)$ and the cumulative (or integrated) parent distribution is given by $\mu(u)=\int_{-\infty}^u \rho(t) {\rm d} t$. The cumulative distribution of the maximum value, $u$ is easy to compute
\be
M_N^{\textrm{max}}(u)={\rm Prob}[u_1 \le u, u_2 \le u,\dots,u_N \le u]=\mu^N(u)\;.
\ee
According to extreme value statistics (EVS) $M_N^{\textrm{max}}(u)$ has a limit distribution for large $N$ and large $x$ in terms of the scaling combination $v=(u-b_N)/a_N$ as
\be
M_N(v)=M_N^{\textrm{max}}(a_N v + b_N) \to M(v)\;,
\ee
and similarly for the extreme density $P_N(v)={\rm d} M_N/{\rm d} v$
\be
P_N(v)=a_N P_N^{\textrm{max}}(a_N v + b_N) \to P(v)\;.
\ee
Here, the $a_N$ and $b_N$ are free up to an additive constant, the value of which is fixed by different standardisation conditions. For the analytical calculation the condition:
\be
M(0)=P(0)=1/e\;,
\label{1st_standardisation}
\ee
is convenient to use, whereas for analysing numerical data it is often better to require
\be
\int_{-\infty}^{\infty} vP(v)=0,\quad \int_{-\infty}^{\infty} v^2 P(v)=1\;,
\label{2nd_standardisation}
\ee
provided the second moment of the distribution exists. In the paper, we refer to Eq.(\ref{1st_standardisation}) and Eq.(\ref{2nd_standardisation}) as \emph{first} and \emph{second standardisation condition}, respectively.

According to EVS theory, the limit distributions can be of three forms, depending on the large $u$ tail of the parent distribution. In a renormalization group treatment\cite{prl,pre} with the first standardisation in Eq.(\ref{1st_standardisation}), the limit distributions are the fixed-point solutions and are given in the form
\beqn
M(v;\gamma)&=&\exp\left[-(1+\gamma v)^{-1/\gamma}\right]\;,\\
P(v;\gamma)&=&(1+\gamma v)^{-1/\gamma-1}\exp\left[-(1+\gamma v)^{-1/\gamma}\right]\;,
\label{eq:M_Fr_v}
\eeqn
for $1+\gamma v \ge 0$ and the different universality classes depend on the value of $\gamma$. For $\gamma>0$ it corresponds to a parent distribution
\be
\mu(u) \approx 1-Au^{-1/\gamma}, \quad u \gg 1\;,
\label{eq:parent}
\ee
which represents the Fr\'echet universality class. For $\gamma=0$ the asymptotic approach of the cumulative parent distribution is faster than a power and represents the Gumbel universality class. Finally, for $\gamma<0$ the asymptotic value is approached at a finite upper border with a power $-1/\gamma$ and is given as the Weibull distribution.

Finite-size corrections in the RG treatment are given in the form
\beqn
M_N(v) &\approx& M(v;\gamma)+\varepsilon_N M_1(v;\gamma,\gamma')\nonumber\\
&=& M(v;\gamma)+\varepsilon_N P(v;\gamma) \psi(v;\gamma,\gamma')\;,
\label{eq:M_Fr_v1}
\eeqn
where
\be
\varepsilon_N=\gamma-\gamma_N \propto N^{\gamma'}\;,
\label{eq:eps_N}
\ee
and
\be
\psi(v;\gamma,\gamma')=\frac{(1+\gamma v)+\gamma'v-(1+\gamma v)^{\gamma'/\gamma+1}}{\gamma'(\gamma'+\gamma)}\;.
\label{eq:psi_v}
\ee
Consequently, the correction term to the cumulative distribution $M_1(v;\gamma,\gamma')$ and that of the density,
$P_1(v;\gamma,\gamma')={\rm d} M_1(v;\gamma,\gamma')/{\rm d} v$ depend on the decay exponent of the parent distribution $\gamma$, as well as the correction exponent $\gamma'$.

\begin{figure}[h!]
\begin{center}
\hskip 1cm
\includegraphics[height=0.7\linewidth]{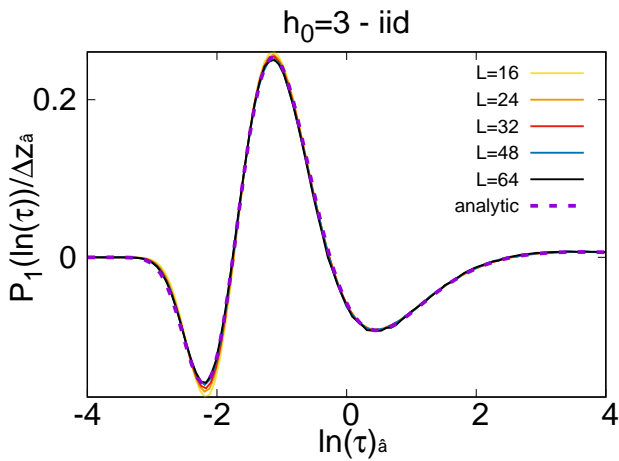}
\end{center}
\caption{\label{fig:pure_power}(Color online) Finite-size corrections to the distributions of the log-extremes for a pure power parent distribution with the second standardisation condition, see text.}
\end{figure}

In practical analysis of the data with a parent distribution in Eq.(\ref{eq:parent}), it is convenient to use logarithmic variables, $x=\ln u$, so that $e^{x-x_0}=1+\gamma v$. Since the parent distribution of $x$ is exponential, the extreme distribution is given by the Gumbel form
\be
\tilde{M}(x,\gamma)=\exp\left(-\exp\left(\frac{-x+x_0}{\gamma}\right) \right)\;,
\label{eq:M_Fr_x}
\ee
and similarly for the probability density
\be
\tilde{P}(x,\gamma)=\frac{1}{\gamma}\exp\left(\frac{-x+x_0}{\gamma}\right)\exp\left(-\exp\left(\frac{-x+x_0}{\gamma}\right) \right)\;,
\label{eq:Frechet}
\ee
so that $\tilde{M}(x_0,\gamma)=1/e$ and $\tilde{P}(x_0,\gamma)=1/(e\gamma)$. In this case, the finite-size correction term is given by $\tilde{P}_1(x;0,\gamma')$.
In the numerical examples, we have $\gamma=0$ and $\gamma'=-1$, and the correction to the density in the second standardisation condition is given by\cite{ozogany}
\beqn
\tilde{P}_1(x;0,-1)&=ae^{-\tilde{x}-e^{-\tilde{x}}}\left[(e^{-\tilde{x}}-1)(1-x/a-e^{-\tilde{x}}) \right. \nonumber \\
&\left. -a^{-2}+e^{-\tilde{x}}\right]\;,
\label{P_1_exp}
\eeqn
with $\tilde{x}=ax+b$, $a=\pi/\sqrt{6}$ and $b=\gamma_E=0.5772156649$ being the Euler-Mascheroni constant. As an illustration, we consider a pure power parent distribution $\mu(u)=1-u^{-1/\gamma},~ u\ge 1$ and in Fig.\ref{fig:pure_power} show the numerically calculated finite-size corrections to the distributions of the log-extremes, denoted by $\ln(\tau)$, for $L=16,24,32,48$ and $64$ with $\gamma=z(h_0=3)$ in the second standardisation condition. These are to be compared with the analytical result in Eq.(\ref{P_1_exp}), since in this case $\gamma'=-1$. The agreement is almost perfect.

\end{document}